\documentclass[12pt, preprint]{aastex}

\shorttitle{Accretion of terrestrial planets}
\shortauthors{Morishima et al.}

\begin{document}


\title{Formation and accretion history of terrestrial planets from runaway growth through to late time 
: implications for orbital eccentricity}



\author{RYUJI MORISHIMA\altaffilmark{1}, MAX W. SCHMIDT}
\affil{Institute for Mineralogy and    
Petrography, Swiss Federal Institute of Technology,
Clausiusstrasse 25, 8092 Zurich, Switzerland}
\email{Ryuji.Morishima@erdw.ethz.ch}

\and

\author{JOACHIM STADEL, BEN MOORE}
\affil{Institute for theoretical physics, University of   
 Zurich, Winterthurerstrasse 190, 8057 Zurich,    
 Switzerland
}


\altaffiltext{1}{The office is at
Institute for theoretical physics, University of   
Zurich, Winterthurerstrasse 190, 8057 Zurich,    
Switzerland}


\begin{abstract}
Remnant planetesimals might have played an important 
role in reducing the orbital eccentricities of the 
terrestrial planets after their formation via giant impacts. 
However, the population and the size distribution of 
remnant planetesimals during and after the giant impact stage
are unknown, because simulations of planetary 
accretion in the runaway growth and giant impact stages have been 
conducted independently.  
Here we report results of direct N-body simulations of the
formation of terrestrial planets beginning with a compact 
planetesimal disk. The initial planetesimal disk has 
a total mass and angular momentum as observed for the
terrestrial planets, and we vary the width (0.3 and 0.5AU) and 
the number of planetesimals (1000-5000). This initial configuration
generally gives rise to three final planets of similar size, and 
sometimes a fourth small planet forms near the location of Mars. 
Since a sufficient number of planetesimals remains, 
even after the giant impact phase, the final orbital
eccentricities are as small as those of the Earth and Venus.

\end{abstract}



\section{Introduction}

The rocky planets are believed to have formed via the accretion of small planetesimals.
The formation mechanism of planetesimals and their initial mass and 
spatial distribution are however still controversial. 
The standard picture of accretion of terrestrial 
planets from planetesimals is as follows.
During the early stages of planetesimal accretion, larger planetesimals 
grow faster than smaller ones owing to their stronger gravity 
\citep{Gre78, WS89}. 
Through this runaway growth stage, 
a few tens of Mars-size protoplanets form with mutual separation of 
10 Hill radii \citep{KI98, KI02}. 
The growth time scale of protoplanets 
is estimated to be  $\sim$ 0.1--1Myr around 1AU, and is longer 
with larger distance from the Sun 
\citep{WS93, KI02}. 
As long as some amount of nebular gas and/or planetesimals remain, 
its damping effect stabilizes the orbits of protoplanets, preventing 
mutual collisions 
\citep{Iwa02, KomI02}.
As the amount of remnant gas and/or planetesimals decreases,
the orbital eccentricities of protoplanets increases due to 
their mutual interactions. Eventually their orbits become chaotic  
and late time giant impacts occur.
During this giant impact stage, whose time scale  
is considered to be $\sim$ 100Myr, the current terrestrial planets form
\citep[e.g.][]{CW98, Agn99, Kok06}.
The orbital eccentricities of planets immediately after 
giant impacts are likely to be much larger than 
those of the current terrestrial planets. 
Therefore, interactions with remnant gas and/or planetesimals is expected to
reduce their eccentricities. 
Reviews for the processes described above 
are given by \citet{Cha04,Nag07}.

Whether remnant gas or remnant planetesimals is more 
important for reducing eccentricities primarily depends 
on the time scale of gas dissipation.   
If the time scale is long, the velocity dispersion of planetesimals 
is suppressed by the gas drag.
Hence, the gravitational focusing effect of protoplanets 
is enhanced, resulting in a fast clean up of remnant 
planetesimals and a lower eccentricity distribution of
the final planets
\citep{AW02, KomI02, KomI04, Nag05, Ogi07}.
On the other hand, if the time scale of gas dissipation 
is short, planetesimals remain unaccreted by protoplanets for a longer
period of time. In this case, remnant planetesimals may be required
to reduce
the eccentricities \citep{Cha01, Obr06, Ray06}.
In this paper, we examine the latter scenario. 
In other words, we ignore the effects of the gas drag and 
the tidal interaction between a gas disk and protoplanets.      
The effects of gas will be investigated in a future study.

There have been several attempts to examine the effect of remnant 
planetesimals based on direct $N$-body simulations \citep{Cha01, Obr06, Ray06},
and with simulations 
using a hybrid-code \citep{KB06}. Direct $N$-body simulations usually adopt
lunar to Mars size protoplanets surrounded 
by smaller planetesimals as initial conditions. 
Direct $N$-body simulations suggest that the eccentricities of final planets 
are further reduced as the total mass of planetesimals increases.
Even for the same total mass of planetesimals, 
the damping effect is stronger with a larger number of smaller planetesimals 
\citep{Obr06, Ray06}. 
However, the total mass 
and mass distribution of the remaining planetesimals are unknown, 
unless mass evolution in the runaway stage is followed. 

The hybrid code of \citet{KB06, BK06},
is able to follow planetary accretion through both the runaway 
and giant impact stages.
In their code,  the mass and velocity distributions of planetesimals 
contained in multi annuli are solved by 
a statistical approach
whereas orbits of protoplanets are solved using
direct $N$-body calculations, that include the effect of 
interaction between planetesimals and protoplanets.
Statistical approaches based on the local approximation
produce consistent results with those obtained from 
direct $N$-body simulations in the runaway stage
\citep[e.g.][]{Ina01}.
However, it is questionable if statistical approaches 
can accurately follow the late 
stage of planetary accretion because orbital eccentricities of remnant
planetesimals are usually very large.
 
Here we report results of direct $N$-body simulations beginning with 
a planetesimal disk until the end of planetary accretion in the 
terrestrial region. We consider compact planetesimal disks 
(initial disk widths of $\le 0.5$AU), 
whose total masses and total angular momenta 
are the same as those of the present terrestrial planets.   
These initial conditions are used since 
accretion simulations beginning from compact disks are 
usually computationally less expensive than those from 
wider disks.  This is the case even with the same initial number of particles, 
because of the rapid decrease in the number of particles through accretion.   
Another reason to adopt compact disks is that total angular momenta of 
final planetary systems obtained from most of previous simulations are 
much larger than for the terrestrial planets, as 
these simulations usually have a super-massive Mars 
\citep[e.g.][]{Cha01, Ray06}. 
This excess angular momentum is likely due to 
initially extended disks. Though Jupiter removes angular momentum, 
mostly from the asteroid region, its effect does not seem to be sufficiently strong
in the terrestrial region. 
As one possibility for this issue, we consider initially compact disks,
supposing that they result from, for example, 
dust migration due to the gas drag prior to formation
of planetesimals \citep[e.g.][]{YS02}.

In \S~2, we explain the numerical methods used in this study.  
We show results of simulations in \S~3. 
We compare our results with previous simulations in \S~4. 
In \S~5, we give some physical interpretations for our simulation results 
using analytic estimations.
We summarize our results in \S~6.

\section{Methods}

The runaway growth stage is shorter than the giant impact stage, but orbits of
a large number of bodies need to be followed. On the other hand, although
fewer bodies are necessary 
for the giant impact stage (unless the effect of fragmentation is considered), 
more care must be taken to accurately follow the orbital evolution over many more
dynamical times.
Taking these physically different types of the accretion stages into account, 
we apply different $N$-body codes to these two stages of the evolution.  

The runaway growth stage is simulated with the parallel 
tree-code {\it PKDGRAV} \citep{Ric00, Sta01} 
for $10^5$ yr with adopting artificially enhanced radii. 
The code uses a fourth-order 
multi-pole expansion for the force calculations, and 
a second-order leap-frog scheme is used for time integration. 
We apply a hierarchical time stepping with the largest time step of 
1.8 days (0.005 yr).  
The opening angle of 0.5 is used as a criterion for searching down the tree.
The energy error in the runaway stage is 
$|\Delta E/E| \sim 10^{-4}$--$10^{-3}$, arising entirely from the integrator.
The error due to the force
calculation using our tree method is negligibly small.    
Using the output of the runaway stage as the initial condition, we simulate
the giant impact stage with the hybrid symplectic code 
{\it Mercury} \citep{Cha99} 
for $2 \times 10^8$ yr without any enhancement of radii. 
This code uses a mixed variable symplectic (MVS) method 
\citep{Kino91,WH91,ST92} for 
orbits around the Sun whereas close encounters are integrated by 
the Bulirsch-Stoer method. We use a fixed time step of 6 days, which is 
the same as or similar to those adopted in the previous works using  
{\it Mercury} \citep{Cha01, Ray06}.
The energy error in the giant impact stage is $|\Delta E/E| \sim 10^{-5}$.
It usually takes less than one computer day  
for a simulation of 
the runaway stage with {\it PKDGRAV}, 
whereas it can take several months to compute
the giant impact stage with {\it Mercury}. 

Whereas MVS type integrators can take much 
larger time steps than those used with the leap-frog scheme, 
the {\it Mercury} code uses direct summation for calculations of the mutual 
gravity force.
{\it PKDGRAV} is thus faster than {\it Mercury} as long as the number of 
particles is larger than several hundred.  
Some comparisons between these two codes are also found 
in \citet{Ray05}.  

The enhancement of radii in the runaway stage is used in order to 
reduce computational time; we use an enhancement factor of radii $g = 4.3$. 
This gives an analogous effect of the gas drag, 
and the growth time scale of protoplanets is 
reduced by a factor of $\sim g^{2}$ 
(Kokubo \& Ida 1996, 2002, see also eq.~[\ref{eq:tgrow}]).
On the other hand, the growth time scale of protoplanets is 
actually reduced by a factor of $\sim \beta^{2}$ by the gas drag,
where $\beta$ is the factor for reduction of planetesimal eccentricities \citep{KI00}. 
Thus, our simulations approximately mimic 
a situation in which the gas disappears suddenly at 
$(g/\beta)^2 \times 10^5$ yr.
This time scale is probably shorter than the typical 
life time of circumsteller disks, $\sim$ a few Myr 
\citep{Hai01}, although the 
exact time scale for formation of planetesimals from dusty gaseous disks is not known.
As long as the number of planetesimals is sufficient, the orbits of protoplanets 
are stabilized during the runaway stage by dynamical friction such that the growth mode is 
not affected by the enhancement of radii \citep{KI96}. 
However, in the transition from the runaway growth 
stage to the giant impact stage, faster clean up of 
remnant planetesimals, due to this approach, usually 
causes final planets to have higher eccentricities.
Thus the radii of particles should be set to realistic values
before planetesimals are too depleted in order to accurately 
examine the effect of remnant planetesimals.  
Additional simulations and analytic calculations were performed to assess the sensitivity of 
the results to varying the time at which $g$ is reduced to unity. 

We use 10 different initial conditions which are summarized in Table~1.
The total mass and angular momentum are assumed to be the same as for  
the present terrestrial planets 
(1.98$M_{\rm E}$ and 1.86$M_{\rm E}$ AU$^{1/2}$ $\sqrt{GM_{\odot}}$, 
respectively, where $M_{\rm E}$, $G$, and $M_{\odot}$ stand for 
the mass of Earth, the gravitational constant, and the solar mass,
 respectively) 
with the central star's mass equal to the solar mass.
The initial width of a planetesimals disk $\Delta_{\rm disk}$ 
is taken to be 0.3 AU or 0.5AU.
The planetesimal mass is assumed to be identical and the 
number of planetesimals $N$ varies from 1000 to 5000. 
The physical density of all the bodies is assumed to be 
$\rho = 2$g cm$^{-3}$.
The surface number density $n(a)$ as a function of the semimajor axis $a$ is 
given by a power law $n(a) \propto a^{\alpha}$ with $\alpha = -1$ or $-2$.
We also conduct two additional simulations for $N=1000$, 
where we switch the code and $g$ at $5 \times 10^4$ yr in order to check 
whether the outcomes are affected by this timing.

It would be very interesting to investigate the accretion of planets using
more extended disks ($\Delta_{\rm disk} > 0.5$AU), but computationally too expensive
with our current codes. 
The rate at which planetesimals merge is slower in a wider disk, particularly at
its extremities,
so we need to use the tree method for the gravity calculation
for longer period of time.
On the other hand, it is not appropriate to use the leap-frog integrator
for the long-term orbital evolution for the following reasons.  
Firstly, since the leap-frog integrator causes a secular error in the longitude of the 
perihelion \citep{KM04},  it does not accurately treat long-term secular
interactions. Secondly, since the standard (or explicit) block multi- timestep
algorithm used for the leap-frog integrator is not time-symmetric, 
the error in the energy accumulates with close encounters \citep{HMM95}.
This is also the case for the higher order Hermite-integrator \citep{KM04}.
We ensure that the transition from {\it PKDGRAV} is chosen conservatively, thus we
achieve high energy conservation as mentioned above.

Although the implicit block time-step algorithm can avoid this problem
\citep{MHKS06}, MVS integrators have considerable advantages for 
simulations of planetary accretion. 
We are therefore implementing the SyMBA integrator \citep{DLL98}
into the latest version of {\it PKDGRAV}, which enables us to simulate 
planetary accretion in wider disks. These simulation results will be reported later.

\section{Results}

\subsection{An example of evolution: Run~6}

As an example, we first explain time evolution of Run~6.
For this simulation the width of the initial disk $\Delta_{\rm disk}$ is
$0.5$AU and the initial number of 
planetesimals $N$ is $3000$.
Figure~1--3 show time evolution of this simulation: 
snapshots on the plane of the semimajor axis versus the orbital eccentricity 
(Fig.~1),
the cumulative number of planetesimals plotted against the mass (Fig.~2), 
and the epicyclic velocity plotted against the mass (Fig.~3).
For detailed analysis, we  
divide the accretional evolution into 
four different stages (the runaway, oligarchic, giant impact, 
and post giant impact stages), rather than two main stages discussed so far.

\subsubsection{Runaway growth stage ($\sim 10^{4}$ yr)}

In the early stage, most of the mass of the system 
is contained in smallest planetesimals. 
In this case the epicyclic velocity,  
$v = \langle a\Omega\sqrt{e^2+i^2}\rangle$ 
(where $\Omega$ is the orbital frequency, and 
$e$ and $i$ are the orbital eccentricity and inclination of a planetesimal, 
respectively),
is regulated by the smallest planetesimals and 
is typically as large as their escape velocity; 
$v_{\rm esc,0} = \sqrt{2Gm_0/(gr_0)}$ (Fig.~3), where $m_0$ and $r_0$ are 
the mass and the radius.
If $v$ is much smaller than the escape velocity 
of the largest body $v_{\rm esc,p} = \sqrt{2Gm_{\rm p}/(gr_{\rm p})}$
(where $m_{\rm p}$ and $r_{\rm p}$ are the mass and the 
radius respectively), 
and is a decreasing function with mass as shown in Figure~3, 
then the largest body starts to grow much faster than nearby objects. 
This growth mode is called runaway growth 
\citep{WS89, KI96, Wei97}. 
At $t \sim 10^4$ yr, the power-law index $q$ 
($dN_{\rm c} \propto m^{q} dm$, where $N_{\rm c}$ is the cumulative number
and $m$ is the planetesimal mass) 
is about $-2.7$ in our simulation (Fig.~2). 
This value is close to  $q \simeq -2.5$ obtained in simulations of 
\citet{KI96, KI00} and the analytical estimate $q = -8/3$ by 
\citet{Mak98}.
(\citet{Mak98} assumed complete
energy partitioning ($v \propto m^{-1/2}$) 
in the strong gravitational limit ($v \ll v_{\rm esc, 0}$).
However, the actual velocity distribution is less steep than this
\citep[see also our Fig.~3]{Raf03}. 
If we assume $v \propto m^{\gamma}$ with $\gamma \sim -1/4$, which is 
a rough approximation of Figure~3 at $t \sim 10^4$ yr, 
the formulation of \citet{Mak98} gives $q = -13/6+\gamma \sim -2.4$.)

\subsubsection{Oligarchic growth stage ($\sim 10^{5}$ yr)}

As large bodies grow, their mutual gravitational interactions leave their
orbits separated by 5--10 Hill radius 
\citep[see also our Fig.~5]{KI95, KI98}. 
The Hill radius $r_{\rm H}$ of a planetesimal of mass $m_{\rm p}$ 
is given by
\begin{equation}
r_{\rm H} = ah_{\rm p} = a \left(\frac{2m_{\rm p}}{3M_{\odot}}\right)^{1/3},
\end{equation}
where $a$ is its semimajor axis and $h_{\rm p}$ is the reduced 
Hill radius.
The largest bodies gravitationally 
influence the velocity evolution of all the neighboring planetesimals 
\citep{IM93}, increasing towards the escape velocity of the protoplanet 
$v_{\rm esc,p}$ (see the panel of $t = 10^5$ yr in Fig.~3).
On the other hand, the velocities of protoplanets $v_{\rm p}$ are 
also influenced by the energy partitioning with surrounding 
smaller planetesimals. 
Indeed, the value of $v_{\rm p}$ for Run~6 is quite close to the 
equilibrium value $v_{\rm p,eq}$ $(\sim v_{\rm esc,0})$, 
which is theoretically estimated neglecting 
the mutual perturbations of protoplanets 
(see eq.~[\ref{eq:eeq2}] in \S~5.3).

Since the growth rate of the largest body slows down at the
expense of its nearby neighbors, the largest intermediate 
mass objects begin
catch up with the largest body \citep{IM93}. 
This growth mode is called oligarchic growth  \citep{KI98}.
At $t = 10^5$ yr in Run~6, about half of the total mass 
is contained in the 10 largest oligarchic bodies. 
Since the growth of smaller planetesimals has
substantially stalled, protoplanets start to separate from the 
continuous size distribution.
Therefore, the number of planetesimals decreases mostly by
 accretion onto protoplanets and not by mutual collisions.
Since $v$ for planetesimals is nearly independent of the mass, 
so is their collision probability with protoplanets. 
Hence, the power-law index $q$ for the mass distribution of 
planetesimals does not change from $-2$ after this stage.

\subsubsection{Giant impact stage ($\sim 10^{6}$ yr)}

Without the damping force by remnant planetesimals and/or gas, 
a multiple protoplanet system undergoes an orbital instability after 
a certain time $T_{\rm inst}$. 
This instability time  $T_{\rm inst}$ depends 
on the orbital separation,  eccentricities, and  
absolute mass of protoplanets 
\citep{Cha96, IT99, Yos99, IO06}.
We switch the code and 
reduce the value of $g$ from 4.3 to 1 at $t = 10^5$ yr.
For Run~6, $T_{\rm inst}$ this time is estimated to be 
$10^5 $--$10^6$ yr from above studies.
With a decreasing total mass of planetesimals that have higher velocities,
the damping due to the dynamical friction of planetesimals
becomes less effective. At this point the orbital instability and mutual collisions of 
protoplanets start to take place.  
We find that the orbital instability starts immediately after $t = 10^5$ yr,
and the number of protoplanets
decreases from 12 at $t = 10^5$ yr to 10 at $t = 2\times 10^5$ yr
(here we assume a protoplanet to be a body with mass 
$> 2\times 10^{26}$g $\sim 50m_0$).

During the giant impact stage, the mass distribution changes mainly 
owing to collisions between protoplanets, 
while the population of small planetesimals does not change so much. 
This can be seen in Figure~2; 
from $t = 10^5$ yr to $t = 10^6$ yr, 
when the number of protoplanets reduces from 
12 to 5, whereas the total number of particles reduces 
only from $\sim 400$ to $\sim 300$. 
Because the velocities of protoplanets are much smaller than those of 
planetesimals, mutual collisions of protoplanets
occur quickly. This is similar to results of simulations
including the damping force due to the tidal interaction 
between a gas disk and protoplanets; \citet{KomI02}
find that the giant impact stage becomes shorter with 
the stronger damping force.

\subsubsection{Post giant impact stage ($> 10^{7}$ yr)}

The number of protoplanets further reduces to 3 at $t = 10^7$ yr,
after two final giant impacts that occur shortly before $t = 10^7$ yr.
Through the giant impact stage, the mutual separation 
between protoplanets normalized by their Hill radii
increases to $\sim 30$. 
The mass distribution becomes completely bimodal (Fig.~2) with
the masses of protoplanets smaller than the isolation mass 
by a factor of $\sim 3$.
The isolation mass is the total mass contained in a ring of width 30 Hill radius 
with the initial surface density (see eq.~[\ref{eq:mp}]). 
This deviation likely  
comes from decrease of the surface density (by a factor of $\sim 2$)
due to expansion of the disk from its initial diameter 
via gravitational scattering of protoplanets.
For remnant planetesimals, the power-law index
$q$ remains to be $\sim -2$ 
and the largest mass is $\sim 50m_0$, which is 
similar to the protoplanet's mass during the runaway to
oligarchic stages.

Since the mutual interactions between protoplanets after the giant impact stage is 
rather weak, the eccentricities of protoplanets are expected to be determined by 
the energy partitioning with remnant planetesimals.
As we discussed for the oligarchic stage,   
$v_{\rm p}$ in the giant impact phase is surprisingly close to the equilibrium value
($\simeq  v_{\rm esc,0}$; see eq.~[\ref{eq:eeq2}]), although some amount of remnant
planetesimals may be necessary to achieve full equilibrium.
For this simulation we find that the fraction of the total mass 
contained in planetesimals is 
0.29 and 0.16 at $t = 10^6$ and $10^7$ yr, respectively.
Naively, one might predict that equilibrium occurs once the mass in 
planetesimals is comparable to the total mass of protoplanets (see \S~5.2).
This slight contradiction might mean that the damping 
due to giant impacts themselves or 
some other unknown mechanism, works effectively.  

\subsection{Evolution of orbital spacing and eccentricities}

Here we analyze the evolution of each simulation quantitatively and 
discuss its dependence on the initial conditions.
In order to examine characteristics of the evolution of the largest bodies,
we define a planet (or a protoplanet) as 
a body having mass larger than $2 \times 10^{26}$g.
The physical meaning of this choice is that 
a body larger than this mass regulates the velocity evolution 
of all the neighboring bodies (see \S~5.3). 
In this case the following discussion does not strongly 
depends on the definition of the minimum mass of planets.
We will explore the following four quantities:
[1] the number of planets, $N_{\rm p}$,
[2] the mass fraction of planets compared to the total mass, $f_{\rm p}$ ,
[3] the orbital spacing of planets normalized by the mutual Hill radius, 
$\bar{b}_{\rm p}$,
[4] the eccentricity of planets, $\bar{e}_{\rm p}$,
and
[5] the ratio of the effective mass of planetesimals to the mean mass of
protoplanets
$m_{\rm eff}/\langle m_{\rm p} \rangle$.
The effective mass of planetesimals is defined as 
$m_{\rm eff} = \langle m^2 \rangle/\langle m \rangle$, 
where $ \langle m^2 \rangle$ and $\langle m \rangle$ are the 
mean squared and mean masses of planetesimals, whose masses are
smaller than $2 \times 10^{26}$g. 
The mass ratio $m_{\rm eff}/\langle m_{\rm p} \rangle$ would be 
important for the evolution of $\bar{e}_{\rm p}$, 
as its equilibrium value due to the dynamical friction is given by 
(eq.~[\ref{eq:epeq}] in \S~5.1)
\begin{equation}
e_{\rm p,eq} \simeq \sqrt{\frac{4m_{\rm eff}}{3\langle m_{\rm p} \rangle}} 
\langle e^2 \rangle^{1/2}. \label{eq:eeq}
\end{equation}

We apply the following form for the 
averaged orbital spacing normalized by the mutual Hill radius
\begin{equation}
\bar{b}_{\rm p} =  \sqrt{
\frac{\sum^{N_{\rm p}-1}_{j=1} ((a_{j+1}-a_j)/r_{{\rm H},j})^2 \mu_j} 
{\sum^{N_{\rm p}-1}_{j=1} \mu_j} \label{eq:aveep}
},
\end{equation}
with the reduced mass
\begin{equation}
\mu_j = \frac{m_{{\rm p},j} m_{{\rm p},j+1}}{m_{{\rm p},j}+m_{{\rm p},j+1}},
\end{equation}
and the mutual hill radius 
\begin{equation}
r_{{\rm H},j} = \frac{1}{2}(a_j + a_{j+1})
\left(\frac{m_{{\rm p},j} + m_{{\rm p},j+1}}{3M_{\cdot}}\right)^{1/3}.
\end{equation}
Here  $a_{j}$, $e_{j}$, and  $m_{{\rm p},j}$ are the semimajor axis,
orbital eccentricity, and mass of the protoplanet $j$ 
in the order of semimajor axis, respectively. 
We use the following form for the averaged eccentricity,
which characterizes the energy of epicyclic motion of planets:   
\begin{equation}
\bar{e}_{\rm p} =  \sqrt{
\frac{\sum_{j=1}^{N_{\rm p}} m_{{\rm p},j} e_{j}^2}
{\sum_{j=1}^{N_{\rm p}} m_{{\rm p},j}}},
\end{equation}
where  $e_{j}$ is the orbital eccentricity of the 
protoplanet $j$. The mass weighted 
eccentricity \citep[e.g.][]{BK06, Ray06}
gives a slightly smaller 
value than the above eccentricity.

Figure~4 shows the evolution of these four quantities for the case of 
$\Delta_{\rm disk} = 0.3$AU.
In the third panel from the top, we also plot the orbital instability time as a 
function of the orbital spacing for the two cases of 
$\bar{e}_{\rm p}/\bar{h}_{\rm p} = 2$ and 4, respectively, 
from \citet{Yos99} (see eq.~[\ref{eq:tinst}]). Here 
$\bar{h}_{\rm p} = (2\langle m_{\rm p}\rangle/(3M_{\odot}))^{1/3}$ 
is the averaged reduced hill radius.
These lines indicate the 
stability of multiple protoplanet systems; if the orbital spacing 
is narrower than these lines, orbital instability will occur. 
Several protoplanets form at $\sim 10^4$ yr and the normalized 
orbital spacing $\bar{b}_{\rm p}$ is about 10, as pointed out by 
\citet{KI98}. Since the orbital instability time 
is $10^5$--$10^6$ yr in such systems, giant impacts between 
protoplanets start around that time as we discussed in the previous 
section. The orbital eccentricity prior to the giant impact stage is
$\sim 0.03$ while it increases up to 0.1 during 
the giant impact stage. The corresponding normalized 
eccentricities, $\bar{e}_{\rm p}/\bar{h}_{\rm p}$, are 2--3 and $\sim 10$,
respectively.
Through the giant impact stage,  $\bar{b}_{\rm p}$ increases 
to $20$--30.  As the radial excursion of planets during the 
giant impact stage determines the final separation of planets, 
the relation between $\bar{b}_{\rm p}$ in the final state 
and $\bar{e}_{\rm p}$ during the giant impact stage
can be approximately represented by \citep{KomI02} 
\begin{equation}
\bar{b}_{\rm p}\bar{h}_{\rm p} \sim 2\bar{e}_{\rm p}. \label{eq:dela}
\end{equation}
This is roughly consistent with our simulation results.

During the post giant impact stage, the orbital eccentricities 
are reduced ($\bar{e}_{\rm p}/\bar{h}_{\rm p} \sim 3$) 
and these values have little dependence on initial parameters 
(we discuss the weak dependence in detail in \S~5.3).
Since $\bar{b}_{\rm p}$ is large enough in the final state, 
the mutual interaction of planets is likely to be unimportant. 
In this case, final eccentricities of planets are expected to 
be determined by the energy partitioning with remnant planetesimals.   
The mass ratio $m_{\rm eff}/\langle m_{\rm p} \rangle$ 
decreases nearly monotonically with time and is about 0.01--0.02 in the
final state (the bottom panel of Fig.~4; the decrease of $m_{\rm eff}/\langle m_{\rm p} \rangle$  
is due to the increase of $\langle m_{\rm p} \rangle$, while 
$m_{\rm eff}$ is nearly constant except at very early times).
Since the mean eccentricity of planetesimals is 
$\sim 0.3$ in the post giant impact stage (Fig.~3), the equilibrium 
eccentricity of planets (eq.~[\ref{eq:eeq}]) is estimated to be 0.03--0.05. 
This is almost the same as the values
obtained in our simulations.

While the final giant impact occurs before $t \sim 10^7$ yr 
in most of the runs, it happens at $t \sim 5 \times 10^7$ yr
for Run~1 ($N=1000$).  Because of that impact, 
the orbital spacing for Run~1 becomes even wider.
This unstable behavior likely suggests that the dynamical 
friction for $N=1000$ is less effective as compared with larger
$N$. However, except for this event, the dependence of the evolutions of 
all the quantities shown in Figure~4 on $N$ is very small.  
Switching $g$ in the earlier time does not affect the quantities 
in the final state shown in Figure~4, either, 
although Run~1b is dynamically more excited during 
the giant impact stage. This is because the mass variation of protoplanets
in Run~1b is somewhat large and smaller protoplanets are dynamically 
enhanced by larger protoplanets. 

Figure~5 is the same as Figure~4, but for the case of 
$\Delta_{\rm disk} = 0.5$AU. The evolution of all the quantities are 
very similar to those in Figure~4, although the early evolution is slightly 
slower due to the lower initial surface density. 
The dependence on $N$ is very small here as well as in Figure~4. 
In particular, the final $\bar{b}_{\rm p}$ has converged to $\sim 30$
for all the runs. However, the final eccentricity 
$\bar{e}_{\rm p}$ is much larger
for the case with earlier switching of $g$ (Run~5b).
This is in an opposite sense to our prediction because dynamical 
friction works more effectively with a higher density of planetesimals. 
We will discuss this issue in the next section along with the
final configurations of the systems.
Except for Run~5b, $\bar{e}_{\rm p}$ and
$m_{\rm eff}/\langle m_{\rm p} \rangle$ in Figure~5
are slightly smaller than those in Figure~4. 
The smaller $m_{\rm eff}$ is due to the smaller surface density 
(the relation is roughly given by $m_{\rm eff} \propto \Sigma^{3/2}$, where
$\Sigma$ is the initial surface density of planetesimals; see \S~5.3).

\subsection{Final systems}

Here we present the orbital parameters of all the final systems obtained 
in our simulations. 
Figure~6 shows snapshots of all of our runs on the $a$-$e$ plane at 200Myr.
Also, the number of planets $N_{\rm p}$, 
the averaged eccentricities of planets $\bar{e}_{\rm p}$,
and the angular momentum deficit of planetary systems $S_{\rm d}$ 
 at 200Myr are summarized in Table~1. 
The angular momentum deficit is defined as 
\citep{Las97,Cha01}
\begin{equation}
S_{\rm d} = \frac{ \sum_{j=1}^{N_{\rm p}} m_{{\rm p},j}\sqrt{a_j}
\left(1 - \sqrt{(1-e_{j}^2)}\cos{i_j}\right)}
{\sum_{j=1}^{N_{\rm p}} m_{{\rm p},j}\sqrt{a_j}}, \label{eq:sd}
\end{equation}
where $i_{j}$ is  the orbital inclination of the planet $j$.  
We take 5Myr averages for $\bar{e}_{\rm p}$ and $S_{\rm d}$. 
For the current terrestrial planets,
we take the mean values between the minimum and maximum orbital 
eccentricities and inclinations from 3Myr orbital integrations 
in \citet{Qui91}. 
This operation roughly corresponds to subtracting only 
the free eccentricities, 
provided that the free eccentricities is larger than the 
forced eccentricities due to giant planets 
\citep[see][Chap.~7.4]{MD99}. 
In fact, the minimum eccentricities and inclinations 
obtained in \citet{Qui91} are almost zero except for Mercury. 
This suggests that the free eccentricity (inclination)
and the forced eccentricity  (inclination) are comparable for the current 
terrestrial planets.

We always obtain three similar size planets between
0.5AU and 1.3 AU, except Run~1 has only two planets in this region. 
The orbital spacing between planets are also quite similar.
We find that more than 95$\%$ of the initial mass and angular momentum
are contained in planets at 200Myr in all our simulations.
Both the averaged eccentricities and the angular momentum deficits 
obtained from our simulations are comparable or even smaller than 
those for the current solar system, except for Run~5b and Run~8 (Table~1).  
Except these two runs, both $\bar{e}_{\rm p}$ and $S_{\rm d}$ 
are smaller for $\Delta_{\rm disk} = 0.5$AU than for 
$\Delta_{\rm disk} = 0.3$AU.
There seems to be a weak tendency that $\bar{e}_{\rm p}$ and $S_{\rm d}$
decrease with increasing $N$. These trends are interpreted in terms of 
the effective planetesimal mass $m_{\rm eff}$, 
if the final $\bar{e}_{\rm p}$ is 
determined by the energy partitioning with planetesimals neglecting 
the mutual interaction of planets (\S~5.3). 
 
Differing from other runs, the mutual interaction between the
innermost two planets in Run~5b and Run~8 is important even at the 
end of simulations, as their orbital separation is narrow. 
This seems related to the spatial distribution of planetesimals
during the giant impact stage. In these runs, we find that
two innermost planets quickly sweep out planetesimals in 
the inner region whilst there are still large numbers of planetesimals 
in the outer region. The outer planets with these planetesimals 
tend to push the middle planet inward. As a result, the 
two innermost planets continue interacting without sufficient 
dynamical friction due to surrounding planetesimals. 
Although our statistics is not sufficient, such a difference in the 
inner and outer region would tend to appear 
when $g$ is reduced at earlier times or if the initial 
planetesimal mass were concentrated in the inner region.

This fact seems related to the existence of the small outermost planet.
When we compare simulations Run~1 and Run~1b, the sizes and locations
of two largest planets are very similar. 
While Run~1b has inner and outer small planets, 
Run~1 has only an outer small planet (near 2.1AU).
Similarly, while the sizes and locations of the three largest planets in 
Run~5 and Run~5b are similar, only Run~5b has an outermost planet.
These facts suggest that the
tendency to have small planets in the inner and outer edges
is stronger in simulations with earlier switching of $g$.  
We interpret this as due to a larger amount of planetesimals that 
are scattered inward (outward) at the inner (outer) edge of the disk 
before they are accreted by planets. Small planets form from
these scattered planetesimals.
A similar trend is also found in the simulations starting with 
a stronger gradient of the surface density in the radial direction
(Runs~4 and 8).
In these systems, the inner planets form quickly while large planets 
have not grown in the outer region. Then inner planets gravitationally
scatter planetesimals and small protoplanets outward. 
The orbital eccentricities of protoplanets scattered outward 
are reduced by the dynamical friction of similarly scattered planetesimals.
Eventually, these protoplanets can have stable orbits near the location 
of Mars and slightly grow as they collide with planetesimals.  

\section{Comparison with previous works}

Our simulation results are different from those starting with 
only protoplanets.  
\citet{Kok06} conducted simulations
starting with $\sim 15$ mars-size protoplanets at $0.5<a<1.5$AU and  
found that most of the final mass is contained in the largest two planets. 
Their final orbital eccentricities are usually higher than those for
the current terrestrial planets.
\citet{CW98} also showed similar results to 
\citet{Kok06} for the case without perturbation of 
Jovian planets (their Model~A). They also found that
mass concentration within fewer planets is strengthened by 
the presence of Jovian planets (their Model~B).
The difference between their results and ours suggests 
that the number of final planets increases with a stronger damping force,
which makes the radial excursion of protoplanets and thus the final separation 
between neighboring planets narrower.  
Indeed, the same trend was also found by \citet{KomI02}, 
who examined the effect of damping due to the gas disk on the accretion of 
protoplanets.

In recent direct $N$-body simulations stating with planetary embryos 
with small planetesimals, 
the total mass of planetesimals 
is half or less than half, and the orbital separation of embryos
is equal to or less than 10 in units of the Hill radii
\citep{Cha01, Obr06, Ray06}.  
Since the dynamical friction of surrounding planetesimals is not 
strong enough to suppress the orbital instability with these initial
conditions (see \S~5.2),
giant impacts start immediately before planets substantially grow by 
accretion.
They adopt a nearly identical size distribution of planetesimals, which
in principle, does not change in such enhanced systems.
Therefore, the effective mass of planetesimals after giant impacts occur
is still given by the initial planetesimal mass. 
Thus, the dependence of the equilibrium eccentricity of planets on
the initial planetesimal mass 
($ e_{\rm p,eq} \sim \sqrt{m_0/\langle m_{\rm p} \rangle}
\langle e^2 \rangle^{1/2} $) 
is much stronger than we find.
In some of their simulations, mutual interaction amongst the final planets 
is insignificant and the final eccentricities 
seem to be close to the equilibrium value.  
For example, \citet{Obr06} adopt the initial 
mass of planetesimals to be $m_0 \simeq 1/400M_{\rm E}$. 
In the late stages of their EJS (eccentric orbits of Jupiter and Saturn) 
simulations, most of the mass supplied to terrestrial planets is 
from small planetesimals with very high speed.  
If we convert their impact speed $\sim 20$km s$^{-1}$ 
in the late stage to the eccentricity around 1AU,  
it gives $\langle e^2 \rangle^{1/2} \sim 0.5$.
Supposing that the mean mass of planets is as large as the Earth's mass,
we obtain  $ e_{\rm p,eq} \sim 0.025$, 
which is even smaller than those for the current terrestrial planets, 
and consistent with their results.
Therefore, we predict that the
final eccentricities of planets would be further reduced if they adopted 
smaller planetesimal masses in their simulations.
However, such a small effective mass of planetesimals in the beginning of 
the giant impact stage might be unlikely if we take the growth of planetesimals
in the runaway and oligarchic stages into account.

\citet{KB06} and \citet{BK06} 
conducted planetary accretion simulations starting 
with very small planetesimals ($r_0 = 1$--5km),
using their hybrid code. 
In their runs, simulations 
starting at $0.86$--1.16AU in \citet{KB06}  have 
similar initial conditions to ours, although our 
disks are slightly more massive. 
The evolution of the number of oligarchic bodies 
(with masses $> \sim 10^{25}$--$10^{26}$g in their simulations) 
and their orbital separation
(their Hill parameter almost corresponds to $1/\bar{b}_{\rm p}$) 
are very similar to our results.
However, the final eccentricities of planets is more 
excited in their simulations
(one of their simulations obtained three planets with $e_{\rm p} \sim 0.1$). 
In fact, 
in all of their simulations starting with wider initial disks ($0.4$--2.0AU), 
the final planetary orbits are more eccentric than the current terrestrial
planets \citep[see Table~I of][]{KB06},
and apparently remnant small planetesimals do not contribute to 
damping of eccentricities of planets.
Since we have not conducted simulations starting with wide disks, 
it is not clear for us if their results obtained with  
a hybrid code are consistent with those obtained 
from direct $N$-body simulations.
We are planning to conduct direct $N$-body simulations with initially 
wide disks to clarify this problem.





\section{Analytic estimates}
In this section,
we interpret our simulation results using analytic estimates.

\subsection{Evolution of velocities of planetesimals 
and protoplanets}
First we provide analytic formulation for the  
evolution of velocities of planetesimals and protoplanets 
necessary for subsequent discussions.
Consider a situation in which protoplanets are spatially separated 
but mutually interact due to the distant perturbations and 
each protoplanet is surrounded by a swarm of planetesimals.
We first consider equal-mass planetesimals, then the formulation 
is extended to 
the case of continuous size distribution (protoplanets are always assumed to be 
equal-mass).
The mass, mean square eccentricity, and surface number density of planetesimals
are represented as $m$, $\langle e^2 \rangle$, and $n$, respectively. 
Corresponding characters for protoplanets are 
$m_{\rm p}$, $\langle e_{\rm p}^2 \rangle$, and $n_{\rm p}$, respectively. 
The scattering cross section for planetesimal-planetesimal encounters 
$\sigma^{m-m'}_{\rm sca}$ and that for protoplanet-planetesimal encounters    
$\sigma^{m_{\rm p}-m'}_{\rm sca}$ are given by \citep{IM93}
\begin{equation}
\sigma^{m-m'}_{\rm sca} = C_{e}\left(\frac{G(m+m')}
{(a\Omega)^2 (\langle e^2 \rangle + \langle e'^2 \rangle)}\right)^2, 
\label{eq:mmsca1}
\end{equation}
\begin{equation}
\sigma^{m_{\rm p}-m'}_{\rm sca} = C_{e}\left(\frac{G(m_{\rm p}+m')}
{(a\Omega)^2 (\langle e_{\rm p}^2 \rangle + \langle e'^2 \rangle)}\right)^2,
\label{eq:mpsca1}
\end{equation}
where $C_{e}$ is the numerical factor of $\sim 40$, 
$G$ is the gravitational constant, $a$ is the distance of the system from 
the Sun,
and $\Omega$ is the orbital frequency. In equation~(\ref{eq:mmsca1}), 
we used primed characters, $m'$ and  $\langle e'^2 \rangle$, to distinguish two 
interacting planetesimal groups (the primed values are averaged later). 
Using the scattering cross sections, 
the change rates of $\langle e^2 \rangle$ and $\langle e_{\rm p}^2 \rangle$ are 
given by 
\begin{eqnarray}
\frac{1}{\Omega}\frac{\langle e^2 \rangle}{dt} &=& 
n \left(\frac{m'}{m + m'}\right)^2  \sigma^{m-m'}_{\rm sca}(\langle e^2 \rangle + 
\langle e'^2 \rangle)
+ n_{\rm p} \left(\frac{m_{\rm p}}{m_{\rm p} + m}\right)^2 
 \sigma^{m_{\rm p}-m}_{\rm sca}(\langle e_{\rm p}^2 \rangle + \langle e^2 \rangle)
 \nonumber \\
&=& C_{e} \left(\frac{G}{(a\Omega)^2}\right)^2 
\left(\frac{nm'^2}{\langle e^2 \rangle + \langle e'^2 \rangle} +
\frac{n_{\rm p}m_{\rm p}^2}{\langle e_{\rm p}^2 \rangle + \langle e^2 \rangle} 
\right), \label{eq:mmsca2}\\
\frac{1}{\Omega}\frac{\langle e_{\rm p}^2 \rangle}{dt} 
&=&
\left(\frac{1}{m_{\rm p} + m'}\right)^2 
\left( 4nm'^2 \langle e'^2 \rangle - 3nm'm_{\rm p}\langle e_{\rm p}^2 \rangle + nm'^2
\langle e_{\rm p}^2 \rangle \right)
\sigma^{m_{\rm p}-m'}_{\rm sca} + \frac{\langle e_{\rm p}^2 \rangle}
{\Omega T_{\rm Inst}},  \label{eq:mpsca2}
\end{eqnarray}
where $T_{\rm Inst}$ is the time scale for orbital instability of protoplanets 
(see \S~5.2).
In equation~(\ref{eq:mmsca2}), the first term stands for the viscous stirring due to 
planetesimal-planetesimal encounters while the second term for 
the viscous stirring due to protoplanet-planetesimal encounters 
\citep{IM93}. In equation~(\ref{eq:mpsca2}), the first term 
stands for the sum of viscous stirring and the dynamical 
friction both due to encounters with 
planetesimals \citep{IM92} while the second term approximately 
accounts for 
the enhancement due to distant interactions between protoplanets.
The rate of change of the inclinations are given by similar equations, but here 
we omit them. 

The surface number density per unit mass is given by $dn/dm$.
As in the simplest case, we assume that $\langle e'^2 \rangle$ is 
independent of mass.
In this case, after integration of 
equations~(\ref{eq:mmsca2}) and 
(\ref{eq:mpsca2}) over the range of $m'$,
$nm'$ and $nm'^2$ in these equations can be replaced by 
\begin{equation} 
\int m' dn, = \Sigma_{\rm s}, \hspace{1em}
\int m'^2 dn = \Sigma_{\rm s} m_{\rm eff}, \label{eq:meff} 
\end{equation}
respectively.
Here $\Sigma_{\rm s}$ and 
$m_{\rm eff} = \langle m'^2 \rangle/\langle m' \rangle$ 
are the surface density and effective 
mass of planetesimals,
respectively. 
Using these averaged quantities, we have simple implications from 
equations~(\ref{eq:mmsca2}) and (\ref{eq:mpsca2}).
Equation~(\ref{eq:mmsca2}) suggests that the velocity evolution of 
planetesimals is regulated 
by protoplanet-planetesimal encounters rather than by 
planetesimal-planetesimal encounters if 
\begin{equation} 
n_{\rm p} m_{\rm p}^2 > f_e \Sigma_{\rm s} m_{\rm eff}, \label{eq:pdom}
\end{equation} 
where $f_e \le 1$ is the numerical factor associated with the velocity distribution.
Equation~(\ref{eq:mpsca2}) suggests that the eccentricity of planets is given by
\begin{equation} 
\langle e_{\rm p,eq}^2 \rangle = \frac{4m_{\rm eff}}{3 m_{\rm p}-m_{\rm eff}}
\langle e^2 \rangle, \label{eq:epeq} 
\end{equation}
in the equilibrium state provided that the mutual interaction 
between protoplanets is negligible.

During the giant impact stage, in which orbits of 
protoplanets cross each other,
the second term in the right hand side of equation~(\ref{eq:mpsca2}) 
can be replaced by
the mutual viscous stirring term:
\begin{equation}
\left(\frac{1}{\Omega}\frac{\langle e_{\rm p}^2 \rangle}{dt}\right)_{\rm mut} 
= \frac{1}{2}\sigma^{m_{\rm p}-m_{\rm p}}_{\rm sca} 
\langle e_{\rm p}^2 \rangle,  
\label{eq:mpsca3}
\end{equation}
with the cross section for the mutual scattering
\begin{equation}
\sigma^{m_{\rm p}-m_{\rm p}}_{\rm sca} = C_{e}\left(\frac{Gm_{\rm p}}
{(a\Omega)^2 \langle e_{\rm p}^2 \rangle}\right)^2.
\label{eq:mpsca4}
\end{equation}
The averaged eccentricity of protoplanets during the giant impact stage
is determined by the balance between the mutual viscous stirring and 
the dynamical friction due to surrounding planetesimals.
Assuming $m_{\rm p} \gg m_{\rm eff}$  and 
$m_{\rm p}\langle e_{\rm p}^2 \rangle \gg m_{\rm eff}\langle e^2 \rangle$, we obtain
\begin{equation}
\langle e_{\rm p}^2 \rangle^{1/2}_{\rm GI} = 
\left(\frac{1}{6}\frac{n_{\rm p}m_{\rm p}}{\Sigma_{\rm s}}\right)^{1/4}
\langle e^2 \rangle^{1/2}.
\label{eq:epgi}
\end{equation} 
The same expression is obtained in \citet{Gol04} except for
a factor of order unity. 
Equation~(\ref{eq:epgi}) somewhat overestimates $\langle e_{\rm p}^2 \rangle^{1/2}$ 
during the giant impact stage as compared with those 
obtained from our simulations. This is probably 
because the motion of all the planets are not enhanced simultaneously
in our simulations as some planets are in stable orbits isolated from others. 
 
\subsection{Comparison of time scales and timing of the onset of giant impacts}

Here we discuss how the timing of the onset of the giant impact stage 
is affected by the radius enhancement factor $g$. 
Since the time and distance can be rescaled for 
our $N$-body simulations, $g$ is physically associated with 
the ratio of the physical radius $r$ to the Hill radius $r_{\rm H}$ as
$g \propto r/r_{\rm H} \propto a^{-1}\rho^{-1/3}$.

For simplicity in this section, we call the most massive body in its
feeding zone of width of $b_{\rm p}r_{\rm H}$, a planet 
(note that the definition of planets used in the 
main text follows the discussion in \S~5.3). 
Defining the mass ratio of the planet with the total mass in the 
feeding zone to be $f_{\rm p}$, the planet mass $m_{\rm p}$ 
is given as 
\begin{eqnarray}
m_{\rm p} &=& 2 \pi f_{\rm p} a b_{\rm p} r_{\rm H} \Sigma
= \left(2 \pi f_{\rm p}b_{\rm p}\Sigma \right)^{3/2} a^3 (2/3M_{\odot})^{1/2}
\nonumber \\
&\simeq& 0.06 \left(\frac{b_{\rm p}}{10} \right)^{3/2}
\left(\frac{\Sigma}{20 {\rm \hspace{0.2em} g \hspace{0.2em} cm}}
\right)^{3/2}
\left(\frac{f_{\rm p}}{0.25}
\right)^{3/2}
\left(\frac{a}{1 {\rm AU}} \right)^{3} M_{\rm E}, \label{eq:mp}
\end{eqnarray}
where $\Sigma$ is the surface density of all the bodies 
(thus assumed to be constant regardless of $f_{\rm p}$).
For $f_{\rm p} = 1$, $m_{\rm p}$ corresponds to the so-called isolation
mass \citep[e.g.][]{KI00}. 

Considering a two component system composed of
planets and surrounding planetesimals,
we introduce the following five timescales, which 
characterize the evolution of planets and surrounding planetesimals.
[1] The growth time scale of planet $T_{\rm grow}$ is given by 
\begin{equation}
T_{\rm grow} = \left|\frac{1}{m_{\rm p}} 
\frac{dm_{\rm p}}{dt} \right|^{-1} =
\frac{m_{\rm p}}{(1-f_{\rm p})\Sigma \sigma_{\rm col} \Omega},  
\label{eq:tgrow}
\end{equation}
 with the collisional cross section $\sigma_{\rm col}$
\citep{GL92}:
\begin{equation}
\sigma_{\rm col} = c_{\rm col}(gr_{\rm p})^2
\left(\frac{1}{3}+\frac{1}{\tilde{v}^2} \right),
\end{equation}
where $c_{\rm col}$ is 
a numerical factor of $\sim 8$ 
(we assume $\langle e^2 \rangle^{1/2} = 2\langle i^2 \rangle^{1/2}$)
and $\tilde{v}$ is the averaged velocity of planetesimals 
normalized by the escape velocity of 
the planet.
[2] The depletion time scale of planetesimals $T_{\rm dep}$ is given by 
\begin{equation}
T_{\rm dep} = \left|\frac{1}{(1-f_{\rm p})\Sigma} 
\frac{d(n_{\rm p}m_{\rm p})}{dt} \right|^{-1} =
\frac{m_{\rm p}}{f_{\rm p}\Sigma \sigma_{\rm col} \Omega}, 
\label{eq:tdep}
\end{equation}
where $n_{\rm p} = f_{\rm p}\Sigma/m_{\rm p}$ is the surface number density 
of planets.
[3] The time scale for the evolution of the velocity of planetesimals 
$T_{\rm sca}$ due to gravitational scattering by planets is given by
\begin{equation}
T_{\rm sca} = \left|\frac{1}{\langle e^2 \rangle} 
\frac{d\langle e^2 \rangle}{dt} \right|^{-1} \label{eq:tsca}
= \frac{m_{\rm p}}{f_{\rm p}\Sigma \sigma_{\rm sca} \Omega}. 
\end{equation}
Here the scattering cross section $\sigma_{\rm sca}$ corresponds to 
$\sigma^{m_{\rm p}-m'}_{\rm sca}$ (eq.~[\ref{eq:mpsca1}]) in the 
limit of $m_{\rm p} \gg m'$ and 
$\langle e_{\rm p}^2 \rangle \ll \langle e'^2 \rangle$
\begin{equation}
\sigma_{\rm sca} = 
c_{\rm sca}(gr_{\rm p})^2 \frac{1}{\tilde{v}^4}, \label{eq:sigsca}
\end{equation}
where $c_{\rm sca}$ is a numerical factor of $\sim 16$.
Note that $\sigma_{\rm sca}$ does not directly depend
on $g$. However, as the velocity can be scaled by the escape velocity 
of the planet, which depends on $g$,
$\sigma_{\rm sca}$ indirectly depends on $g$ as represented by 
equation~(\ref{eq:sigsca}).
[4] The time scale for damping the velocity of planets $T_{\rm damp}$ 
due to dynamical friction of surrounding planetesimals is given by 
\begin{equation}
T_{\rm damp} = \left|\frac{1}{\langle e_{\rm p}^2 \rangle} 
\frac{d\langle e_{\rm p}^2 \rangle}{dt} \right|^{-1} \label{eq:tdamp}
= \frac{m_{\rm p}}{(1-f_{\rm p})\Sigma \sigma_{\rm sca} \Omega}. 
\end{equation}
[5] The time scale for the orbital instability $T_{\rm inst}$
of a multiple protoplanetary 
system without any damping force represents 
either the time of the first collision or the first close encounter 
\citep[see][for comparison of 
$T_{\rm inst}$ with different types of definitions]{Yos99}.
The form of  
$T_{\rm inst}$ is empirically given by \citep{Cha96, Yos99, IT99, IO06}
\begin{equation}
\log \left(\frac{T_{\rm inst}}{T_{\rm orb, 1}}\right) = c_1 b_{\rm p} + c_2,
\label{eq:tinst}
\end{equation}
where $T_{\rm orb, 1}$ is the orbital period of 
the innermost protoplanet, and 
$c_1$ and $c_2$ are numerical coefficients. 
These coefficients  
depend strongly on the orbital eccentricity and relatively weakly on  
the absolute averaged mass and the variation of masses.
The dependence of $c_1$ and $c_2$ on orbital eccentricities is 
summarized in Table~III of \citet{Yos99}.
For $b_{\rm p} = 10$ and $\bar{e}_{\rm p} \sim 4\bar{h}_{\rm p}$,
$T_{\rm inst}/T_{\rm orb, 1} \sim 10^5$-- $10^6$. 

For simplicity, we normalize all the timescales as follows:
\begin{equation}
\tilde{T} =  \frac{c_{\rm col} \Sigma (gr_{\rm p})^2}{m_{\rm p}} \Omega T.  
\end{equation}
With this normalization, 
all the time scales except for $\tilde{T}_{\rm inst}$ depend 
only on $\tilde{v}$ and $f_{\rm p}$ as 
\begin{eqnarray}
\tilde{T}_{\rm grow}^{-1} &=& \left(1-f_{\rm p}\right)
\left(\frac{1}{3} + \frac{1}{\tilde{v}^2}\right), \nonumber \\
\tilde{T}_{\rm dep}^{-1} &=& f_{\rm p}
\left(\frac{1}{3} + \frac{1}{\tilde{v}^2}\right), \nonumber \\
\tilde{T}_{\rm sca}^{-1} &=& 2f_{\rm p}
\frac{1}{\tilde{v}^4}, \nonumber \\
\tilde{T}_{\rm damp}^{-1} &=& 2\left(1-f_{\rm p}\right)
\frac{1}{\tilde{v}^4}.
\end{eqnarray}
It should be noted that now the dependence on $g$ 
is included only in $\tilde{T}_{\rm inst}$.

In order to obtain the time scales as functions of $f_{\rm p}$,
we consider the evolution of $\tilde{v}$. 
In the early stages the smallest planetesimals dominate the mass of 
the system so that the velocity is as large as the escape velocity 
of planetesimals.  The exact value of $\tilde{v}$ 
at the initial state (when $m_{\rm p} = m_0$) is determined by 
the balance between the mutual scattering and the collisional damping.  
Since the ratio of these time scales is given by the ratio of 
$\sigma_{\rm col}$ to $\sigma_{\rm sca}$, we obtain  
$\tilde{v} = 1.17$ for the initial state. 
Since the dimensional velocity of planets is $1.17v_{\rm esc,0}$,
$\tilde{v}$ decreases as planets grow (or with increasing $f_{\rm p}$).
When $f_{\rm p}$ reaches to a certain value, 
planets start to regulate the velocities of surrounding planetesimals. 
In this case, $\tilde{v}$ evolves as planets grow. 
Thus, $\tilde{v}$ is obtained from 
$\tilde{T}_{\rm grow} = \tilde{T}_{\rm sca}$ 
\citep{Dai06}. 
The value of $\tilde{v}$ increases with $f_{\rm p}$ to 
1.17 at $f_{\rm p} = 0.5$. 
For $f_{\rm p} > 0.5$, 
$\tilde{T}_{\rm dep}$ becomes shorter than $\tilde{T}_{\rm sca}$,
if $\tilde{v} > 1.17$. 
This means that planetesimals collide with planets 
before their velocities are further enhanced. 
Therefore, $\tilde{v}$ takes a constant 
value, 1.17, for $f_{\rm p} > 0.5$.  

We plot the evolution of $\tilde{v}$ as a function of $f_{\rm p}$ in Figure~7.
In the same figure, we also plot 
$\tilde{T}_{\rm grow}$, $\tilde{T}_{\rm sca}$, $\tilde{T}_{\rm damp}$, and 
$\tilde{T}_{\rm inst}$ with the parameters used in 
Run~6 and $b_{\rm p} = 10$. 
For $\tilde{T}_{\rm inst}$, we  
take $T_{\rm inst}/T_{\rm orb} = 5 \times 10^5$ with 
$g = 1$ and 4.3 as examples. This time scale would be appropriate 
judging from our simulations.
It should be noted that $\tilde{T}_{\rm dep} = \tilde{T}_{\rm sca}$ 
for $f_{\rm p} > 0.5$ and $\tilde{T}_{\rm dep}$ further increases 
with decreasing $f_{\rm p}$ for $f_{\rm p} \le 0.5$, although we do not plot 
$\tilde{T}_{\rm dep}$ on Figure~7 in order to avoid confusion due to 
too many lines. 

\citet{Iwa02} and \citet{IO06}
showed that orbital instability is prevented 
when $\tilde{T}_{\rm damp} < c_3\tilde{T}_{\rm inst}$, 
with a coefficient $c_3$
of the order of unity (we assume $c_3 = 3$  after \citet{IO06}).
Hence, the value of $f_{\rm p}$ at the onset of giant impacts
is estimated from the equation $\tilde{T}_{\rm damp} = 3\tilde{T}_{\rm inst}$,
and is 0.3--0.4 and 0.8--0.9 for $g = 1$ and 4.3, respectively, in Figure~7.
However, Figure~7 also suggests that  
$\tilde{T}_{\rm inst}(g =4.3) > \tilde{T}_{\rm dep}$, 
which means that remnant planetesimals are depleted before 
the giant impact phase starts, as long as we keep $g = 4.3$. 
Complete depletion of planetesimals keeping  
$b_{\rm p} \sim 10$ is found in simulations with $g = 6$ in 
\citet{KI02} and \citet{LR05}, 
and we also found the same results with additional tests.
Therefore, in order to keep some amount of planetesimals 
at the onset of the giant impact stage, it must start before 
$f_{\rm p}$ reaches 0.5 (strictly speaking, this value is slightly higher than 0.5 for $c_3 > 1$). 
This also corresponds to the condition 
for the damping of enhanced eccentricities of planets during 
the giant impact stage. The condition under which the
eccentricities of planets are damped before depletion of 
planetesimals is given by $\tilde{T}_{\rm damp} < \tilde{T}_{\rm dep}$, 
which gives $f_{\rm p} < 0.5$ (Fig.~7). 
For the case of $g = 1$, this condition is satisfied. 
The threshold value of $g$ to satisfy the condition is roughly estimated 
to be 3.

For our simulations (except Runs~1b and 5b), 
we reduced $g$ from 4.3 to 1 when $f_{\rm p}$ is about 0.6 (Figs.~4 and 5).
After reducing $g$, $\tilde{T}_{\rm inst}$ becomes smaller than 
any other timescale (Fig.~7). 
Thus, the giant impact phase rapidly begins
(the instantaneous reduction of $g$ causes drop of $\tilde{v}$ to 
$\sim 1.17/\sqrt{4.3}$, but $\tilde{T}_{\rm sca}$ is shorter than 
$\tilde{T}_{\rm grow}$ by a factor of 4.3 with this small $\tilde{v}$
and thus $\tilde{v}$ increases near to 1.17 again before 
$f_{\rm p}$ substantially increases).
On the other hand, for Runs~1b and 5b, 
$f_{\rm p}$ is about $0.4$ when we reduce $g$ at $5 \times 10^4$ yr.
In these cases giant impacts do not start immediately as 
 $\tilde{T}_{\rm damp} \sim 3\tilde{T}_{\rm inst}$ at $f_{\rm p} \sim 0.4$
(this can be seen in Figs.~4 and 5 as $N_{\rm p}$ does not change 
immediately after reducing $g$). 
In fact, after $f_{\rm p}$ increases to $\sim 0.5$, giant impacts start
in these simulations.
Therefore, it is expected that reducing $g$ at an earlier time 
does not affect the results (we have now conducted some simulations 
keeping fixed $g = 1$ and obtained consistent results; 
these results will be reported in another paper).

To summarize, for simulations with a constant $g$ throughout the entire
accretion stage, sweeping up all the planetesimals and 
a subsequent giant impact stage are expected for $g > 3$, 
whereas giant impacts during accretion of planetesimals
and a subsequent damping of eccentricities of planets due to 
the dynamical friction are expected for $g < 3$.
For $g < 3$, the evolution and final state of a system are 
expected to weakly depend on $g$, from the comparison between 
Runs~1 and 1b and that between Runs~5 and 5b. 

\subsection{Minimum oligarchic mass and equilibrium eccentricity of protoplanets}

Once the mass of the largest body is above a certain critical mass, 
the velocity evolution of 
neighboring planetesimals is primarily regulated by the largest body.
We call this critical mass the minimum oligarchic mass.  
Then, the growth of smaller planetesimals near the largest body 
is stalled and as the largest body grows 
it starts to separate from the continuous size distribution of planetesimals.
Therefore, the planetesimal size distribution after this stage is 
expected to be a continuous distribution with the maximum mass being 
the minimum oligarchic mass. This size distribution determines the equilibrium 
eccentricity of planets after the giant impact stage. 
Here we estimate the planetesimal size distribution, when 
the largest body starts to regulate the velocity evolution of 
all of its neighboring planetesimals.

We consider the power-law size distribution for planetesimals 
$dn \propto m^{q}dm$ with the upper and lower cutoff masses 
$m_{\rm p}$ and $m_0$. The condition that the largest body regulates 
the velocity evolution is again given by (eq.~[\ref{eq:pdom}])
\begin{equation}
m_{\rm p}^2 > f_e m_{\rm T} m_{\rm eff}, \label{eq:pdom2}
\end{equation}
where $f_e$ is a factor associated with the velocity distribution
and is slightly smaller than unity and $m_{\rm T}$ ($= \Sigma_{\rm s}/n_{\rm p}$) 
is the total mass of planetesimals (excluding the largest body or a planet) 
in the heated region, where velocities are regulated 
by a planet \citep[see][]{IM93}, with $n_{\rm p}$ being the 
surface number density of planets (the sizes of the heated region and 
the feeding zones are similar).
We also define the cumulative number in the heated region
to be  $N_{\rm h} = n/n_{\rm p}$, 
and now the size distribution is given by $dN_{\rm h} = k m^{q} dm$.
Using the condition for the mass of the largest planetesimals 
(or the second largest body in the heated region) $m_{\rm L}$
\begin{equation}
2 = \int^{\infty}_{m_{\rm L}} dN_{\rm h} = 2\frac{km_{\rm L}^{q+1}}{-q-1},
 \hspace{1em} ({\rm for} \hspace{0.5em} q < -1)
\end{equation}
the total mass $m_{\rm T}$ and the effective mass $m_{\rm eff}$ are,
 respectively, given by 
\begin{eqnarray}
m_{\rm T} &=& \int^{m_{\rm L}}_{m_0} m dN_{\rm h} = 2\frac{q+1}{q+2} 
\left[\left(\frac{m_{\rm L}}{m_0}\right)^{-q-2} - 1\right] 
m_{\rm L}, \hspace{1em} ({\rm for} \hspace{0.5em} q \ne -2)
\label{eq:mt1}\\
m_{\rm T}m_{\rm eff} &=& \int^{m_{\rm L}}_{m_0} m^2 dN_{\rm h} =
-2\frac{q+1}{q+3} \left[ 1-\left(\frac{m_0}{m_{\rm L}}\right)^{q+3}\right]
m_{\rm L}^2. 
\hspace{1em} 
({\rm for} \hspace{0.5em} q \ne -3) \label{eq:meff1}
\end{eqnarray} 
For $q = -2$, we have  $m_{\rm T} = 2m_{\rm L}\ln(m_{\rm L}/m_0)$. 

Substituting  equation~(\ref{eq:meff1}) and the relation  
$m_{\rm L} = 2^{1/(q+1)}m_{\rm p}$ into equation~(\ref{eq:pdom2}),
the condition approximately becomes  
\begin{equation}
1 < - f_e \frac{q+1}{q+3} 2^{(q+3)/(q+1)} \label{eq:pf},
\end{equation}
which gives $q > -2.2$ for $f_e=1$. 
Since $f_e$ is expected to be slightly smaller than unity, 
$q$ would be slightly smaller than $-2.2$
when the largest body starts to regulate the velocity evolution. 
This might correspond to the value $q\simeq -2.5$ obtained by \cite{KI96}.

The fraction of planetary mass $f_{\rm p}$ is given by  
\begin{equation}
f_{\rm p} = \frac{m_{\rm p}}{\int^{m_{\rm p}}_{m_0} m dN_{\rm h}} = 
\frac{q+1}{q+2}\left[(m_{\rm p}/m_{0})^{-q-2}-1\right]^{-1}. \label{eq:fp2}
\end{equation}
Substituting it 
into equation~(\ref{eq:mp}),  
we obtain the minimum oligarchic mass $m_{\rm p,oli} = m_{\rm p}(q \sim -2.2)$ 
as function of $q$. 
This is shown in Figure~8 as well as the corresponding 
$m_{\rm eff}$ for the parameters of Run~6.  
Using $m_{\rm p}(q)$, we also plot the evolution of 
$q$ as a function of $f_{\rm p}$ in Figure~7. 
As we estimated above,  Figure~7  suggests that 
$\tilde{v}$ starts to increase at $q \sim -2.5$ because of 
the gravitational scattering of planets.
Therefore, when $q \sim -2.5$,  
growth of small planetesimals slows down as well as 
the evolution of their size distribution.

However, at this stage, the velocities are still not high enough to suppress the growth 
of large planetesimals.   
Therfore, mass transfer from small planetesimals to large planetesimals 
further increases $q$ to $\sim -2$.  
When $q \sim -2$, the velocity of planetesimals in the heated region 
is almost regulated by a single planet, whereas
the contribution from a planet and all the other bodies are the same for
$q \sim -2.5$. 
As a result, 
the velocity of planetesimals is enhanced toward the escape velocity of 
the planet. Then, actual separation of the planet
from the continuous size distribution starts as we showed in \S~3.1 
(that is also why we keep $q = -2$ for large $f_{\rm p}$
in Fig.~7, although there is no physical reason for $q$ to be strictly $-2$). 
Therefore, an appropriate 
minimum oligarchic mass to give the size distribution after the giant 
impact stage seems  
to be $m_{\rm p,oli}$ for $q \simeq -2$ 
($m_{\rm p, oli} \sim 60m_0$ for $q=-2$ whereas $\sim 20m_0$ for $q=-2.5$). 
In the case of $q=-2$, 
$m_{\rm p,oli}$ and $m_{\rm eff}$ 
depend on the initial planetesimal mass $m_{0}$
very weakly as $m_{\rm p,oli}(q=-2) \propto [\ln{(m_{\rm p,oli}/m_0)}]^{-3/2}\Sigma^{3/2}$
and $m_{\rm eff}(q=-2)= 0.5m_{\rm p,oli}(q=-2)/\ln{(0.5m_{\rm p,oli}/m_0)}$.

Using the form of $m_{\rm eff}$ for $q = -2$, we rewire 
the equilibrium eccentricity (eq.~[\ref{eq:eeq}]) as 
\begin{equation}
e_{\rm p,eq} = \left(\frac{2}{3}\frac{m_{\rm p,oli}/m_0}{\ln{(0.5m_{\rm p,oli}/m_0)}} \right)^{1/2}
\left(\frac{m_0}{m_{\rm p}}\right)^{1/6} v_{\rm esc,0}, 
\hspace{1em} ({\rm for} \hspace{0.2em} q = -2). \label{eq:eeq2}
\end{equation}
Substituting $m_{\rm p,oli}(q = -2)$, 
which is obtained from eqs.~[\ref{eq:mp}] and [\ref{eq:fp2}], into eq.~[\ref{eq:eeq2}],
we calculate $e_{\rm p,eq}$ for various cases 
of $N$ (or $m_0$) and $\Delta_{\rm disk}$ (or $\Sigma$), assuming three final planets 
(thus $m_{\rm p} = 0.66M_{\rm E}$) and the velocity of planetesimals to be the escape velocity 
of a planet ($v = v_{\rm esc, p}$). 

The calculated values are plotted in Fig.~9 and compared with our $N$-body simulation results. 
We find good agreements between analytic estimates and simulations, 
as long as the planet-planet interactions are not important in the final state.
Since the dependence of $e_{\rm p,eq}$ on $N$ is very weak for $N > 1000$,
we need more runs for statistics.
We also conducted additional simulations for the case of $N = 100$ and results 
are plotted in the same figure. 
In most of simulations with $N = 100$ (we conducted four 
simulations for each $\Delta_{\rm disk}$), 
giant impacts occur after nearly complete 
sweep up of remnant planetesimals. 
Therefore, it would not be appropriate to apply 
our analytic estimate to the case of $N = 100$. 
Nevertheless, we find coincidental good agreements 
between the averaged $e_{\rm p,eq}$'s for simulations with
$N = 100$ and those from the analytic estimates.
 
Finally, let us discuss what will happen if our simulations started with a very small size 
distribution of planetesimals. In Figure~8, we also plot $m_{\rm p,oli}$ and $m_{\rm eff}$
as functions of $q$ for the parameters used in 
\citet{Wei97} ($m_0 = 4.8 \times 10^{18}$g), who conducted planetary accretion 
simulations in the runaway and oligarchic stages with their multi-zone code.  
For $(m_{\rm p,oli}/m_{0})^{-q-2} \gg 1$ (this is not the case for our simulations),
the minimum oligarchic mass is given by 
\begin{equation}
m_{\rm p,oli} = \left(\frac{q+1}{q+2} 2 \pi  m_0^{-q-2} \Sigma a^2 
b_{\rm p} \left(\frac{2}{3M_{\odot}}\right)^{1/3} \right)^{-1/(q+4/3)}.
\end{equation}
This equation indicates that $m_{\rm p,oli}$ decreases rather strongly
with $m_{0}$ for small $q$, as $m_{\rm p,oli} \propto m_{0}^{(q+2)/(q+4/3)}$.
Therefore, the minimum oligarchic mass with $q = -2.5$ for $m_0 = 4.8 \times 10^{18}$g 
is much smaller than ours ($m_0 = 3.94 \times 10^{24}$g), 
and oligarchic bodies started to regulate the velocity evolution even 
when $f_{\rm p}$ is very small (estimated to be $\sim 10^{-3}$). 
On the other hand, $m_{\rm p,oli}$ and  $m_{\rm eff}$ for $q = -2$
are only one order of magnitude smaller than those for Run~6. 
Indeed, in simulations of \citet{Wei97}, 
the maximum mass of remnant planetesimals is $\sim 10^{25}$g
(whereas oligarchic bodies have masses $\sim 10^{27}$g), 
suggesting $q \sim -2$.
Therefore, from eq.~[\ref{eq:eeq2}] the final equilibrium eccentricities is expected to be 
smaller than those in our simulations only by a factor of 3--4.
In the discussion here, we ignored the effect of damping due to 
mutual collisions,
which would reduce the eccentricities of planetesimals.
If the fast clean up of remnant planetesimals happens due to the collisional 
damping, the final planetary system might be unstable, as in the case of a large $g$.  

\section{Conclusions} 

We have conducted direct $N$-body simulations of the
formation of terrestrial planets beginning with a compact 
planetesimal disk, with 
the total mass and angular momentum being those  
for the current terrestrial planets. 
In most of the cases, a planetesimal disk results in 
three planets of similar size, and sometimes a fourth
small planet forms around the location of Mars. 
Since a sufficient number of planetesimals remain 
even after giant impacts of protoplanets, 
orbital eccentricities of the final planets 
are as small as those of the current Earth and Venus.
This is a very nice success of our simulations which test
the standard model for planet formation.

The final eccentricities of planets are nearly
in the equilibrium state for the energy partitioning with 
remnant planetesimals, meaning that the mutual interactions 
of planets in the final state is unimportant in most cases.
The final eccentricities of planets depends on the initial mass 
of planetesimals only very weakly, and on the surface density 
relatively strongly. These dependences are interpreted in terms of 
 the effective mass of remnant planetesimals 
$(m_{\rm eff} = \langle m^2 \rangle/\langle m \rangle)$, which 
determines the strength of the gravitational scattering effects of planetesimals.
The mass distribution of remnant planetesimals is approximately
represented by a power law distribution, $dn \propto m^{q}dm$, with $q \sim -2$,
with the upper cut off mass (we call it the minimum oligarchic mass)
which increases very weakly with the initial mass of planetesimals.
Therefore, the dependence of the effective mass on the  
initial mass of planetesimals is very weak as well. 

In a few of our simulations, planet-planet interactions are important even at the end of 
simulations and the orbits of final planets can be
much more eccentric than for our terrestrial planets. This situation seems to appear when the gradient 
of population of planetesimals in the radial direction is large 
during the giant impact stage. However, the number of our simulations is  
still too small to statistically discuss the conditions that are responsible for
final planetary orbits.

\acknowledgments
We appreciate an anonymous reviewer for useful comments.
We are grateful to Derek Richardson for providing us with his version of 
{\it PKDGRAV}.
We thank Shigeru Ida, Makiko Nagasawa, and Eiichiro Kokubo for fruitful 
discussions.
Our simulations have been conducted with the zBox1 and zBox2 supercomputers 
at the University of Zurich. 
We thank Doug Potter for his management of the computers.

\clearpage
\begin{figure}
\begin{center}
\includegraphics[width=.85\textwidth]{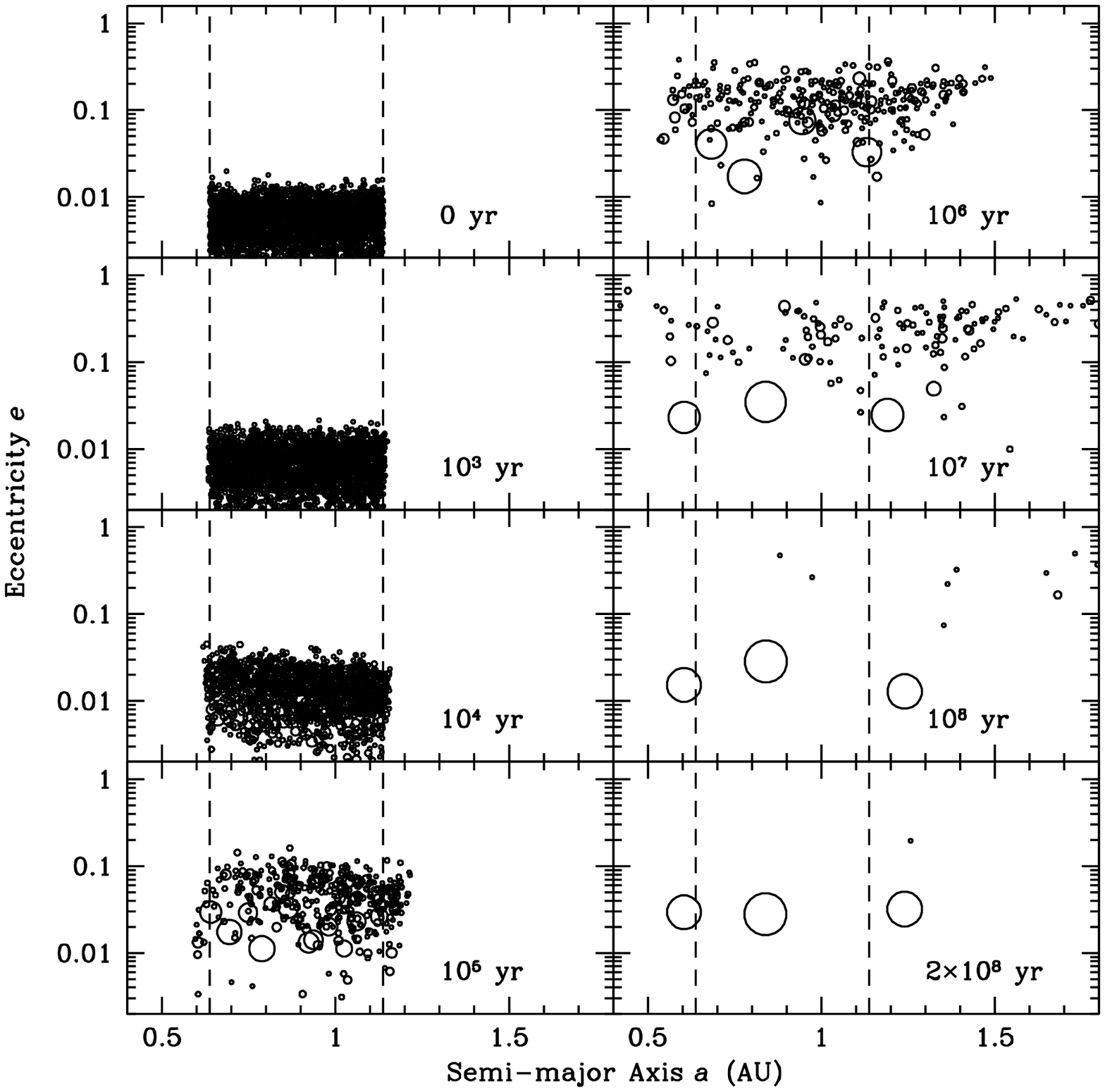}
\end{center}
\caption{Snapshots of Run~6 on the $a$--$e$ plane. 
Vertical dashed lines are the inner and outer edges of 
the initial planetesimal disk. The circles represent planetesimals and planets,
and plotted radius sizes are proportional to 
the actual radii without artifical enhancement.}

\end{figure}

\clearpage

\begin{figure}
\begin{center}
\includegraphics[width=.85\textwidth]{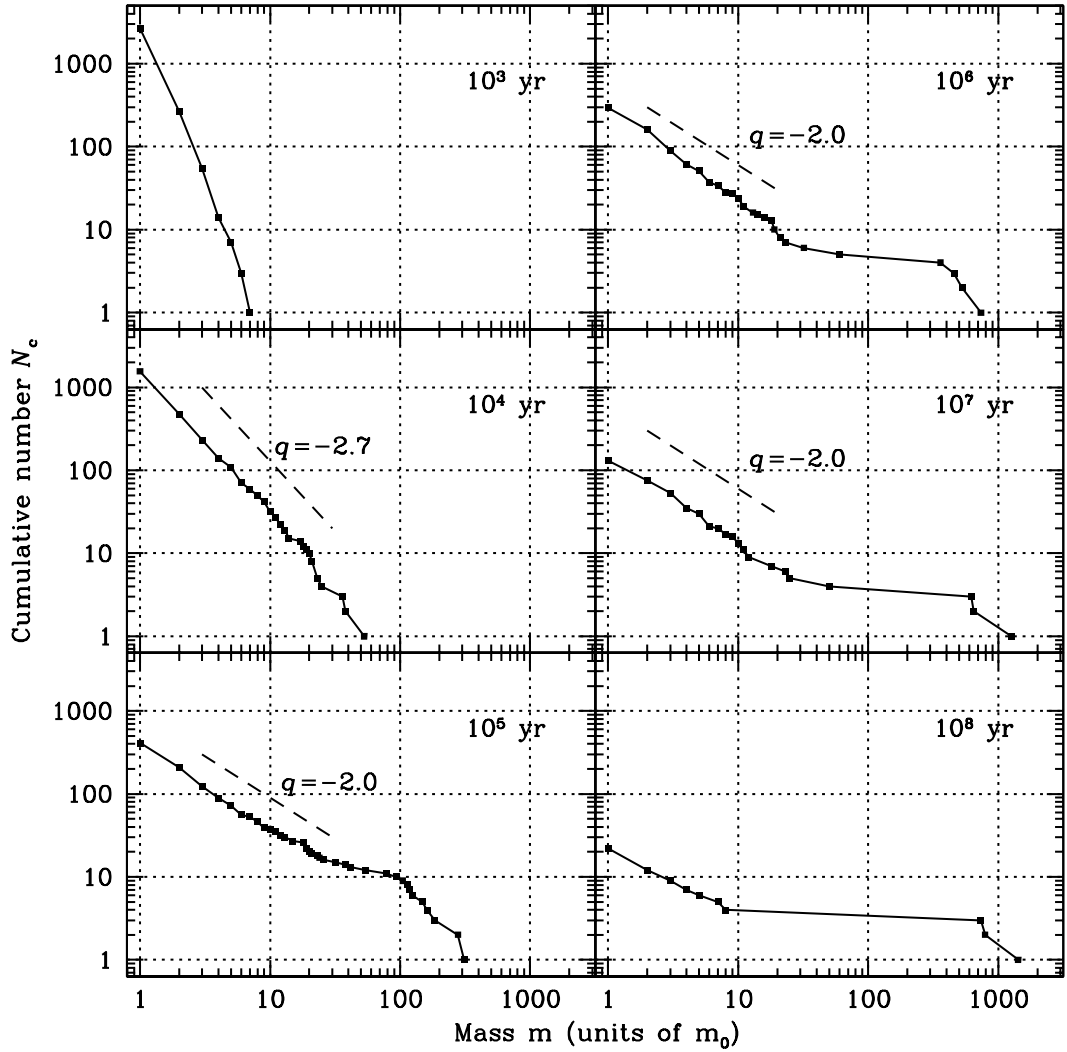}
\end{center}
\caption{Evolution of the cumulative number vs. mass in units 
of the initial mass for Run~6.
The slope of the approximated power-law distribution, 
$dN_{\rm c} \propto m^{q}dm$, is shown in each panel.}
\end{figure}

\clearpage

\begin{figure}
\begin{center}
\includegraphics[width=.85\textwidth]{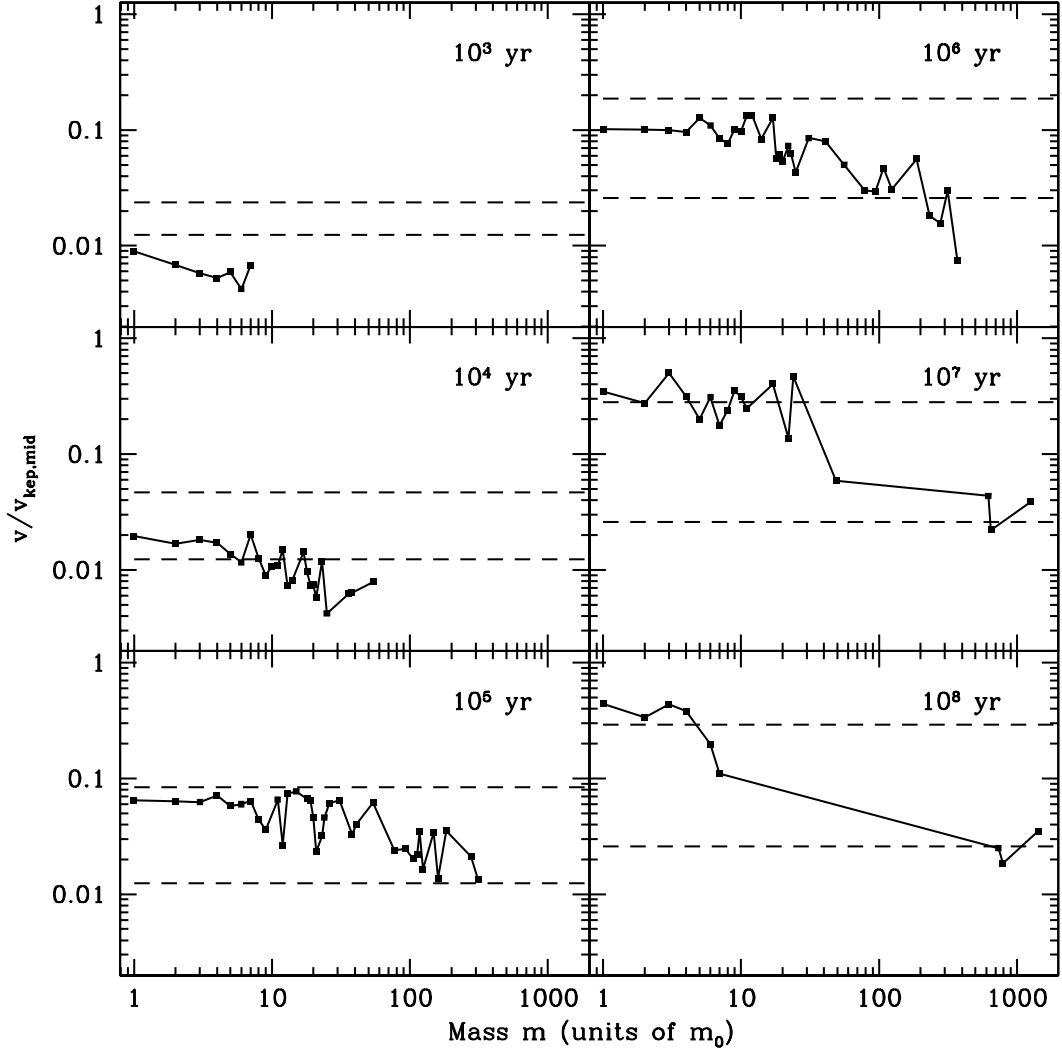}
\end{center}
\caption{Evolution of the epicyclic velocity vs. mass 
for Run~6. The velocity is normalized by the Keplerian velocity 
$v_{\rm kep,mid}$ at $a=0.89$AU.
The upper and lower horizontal dashed lines represent 
the escape velocities of the largest body and the smallest body, 
respectively. Note that the escape velocity of the smallest body 
increases by a factor of $\sqrt{g}$ after $10^5$ years as we reduce the
radius enhancement factor $g$ from 4.3 to 1.}

\end{figure}

\clearpage
\begin{figure}
\begin{center}
\includegraphics[width=.85\textwidth]{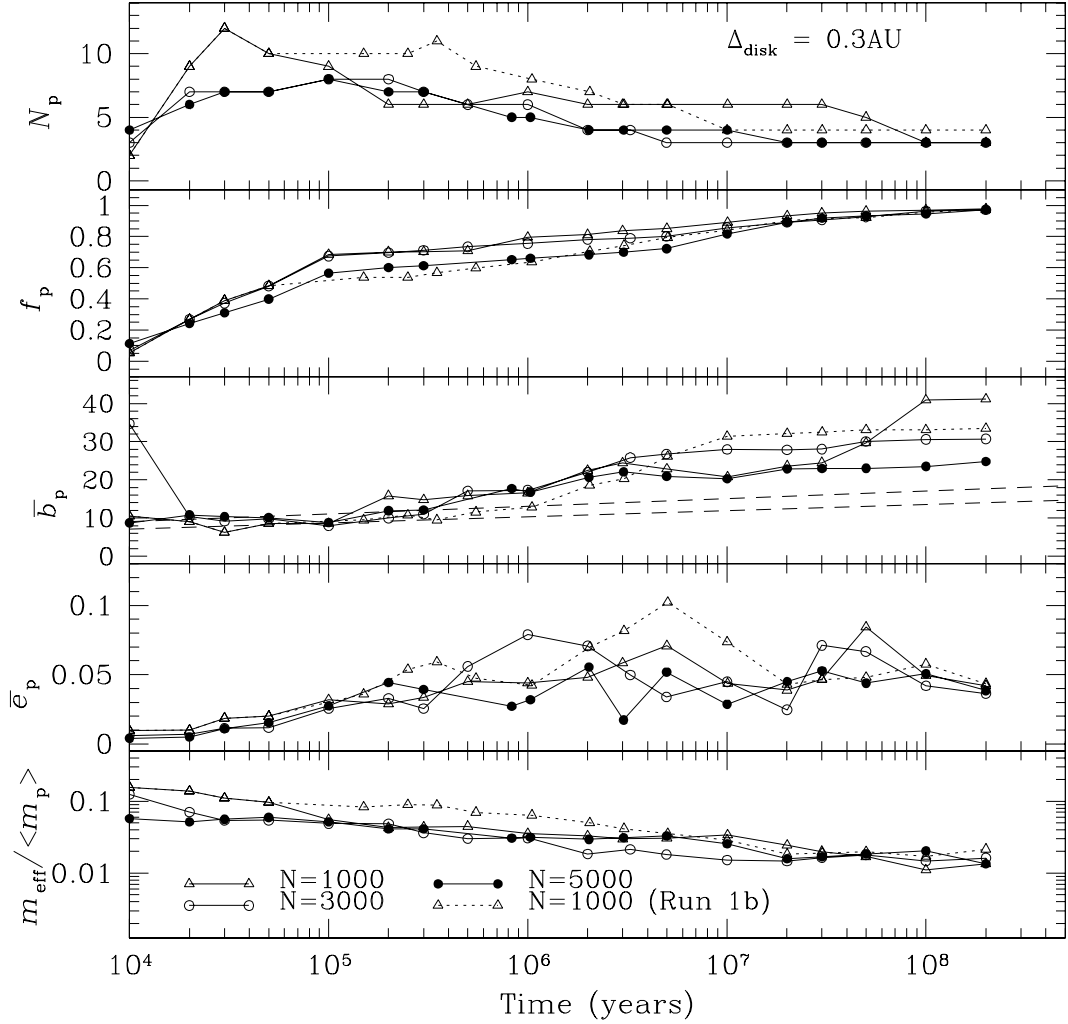}
\end{center}
\caption{Time evolution of the averaged quantities for 
$\Delta_{\rm disk} = 0.3$AU (Runs~1, 1b, 2, and 3).
From top to bottom, the panels show
the number, mass fraction, orbital spacing, and orbital eccentricity of planets 
($m > 2 \times 10^{26}$ g), and the ratio 
of the effective mass of planetesimals to the mean mass of planets.    
In the third panel, the relations between the orbital spacing 
and orbital instability time for $\bar{e}_{\rm p}/\bar{h}_{\rm p} = 2$ and 4 
\citep{Yos99} are shown by lower and upper dashed lines, respectively.}

\end{figure}

\clearpage
\begin{figure}
\begin{center}
\includegraphics[width=.85\textwidth]{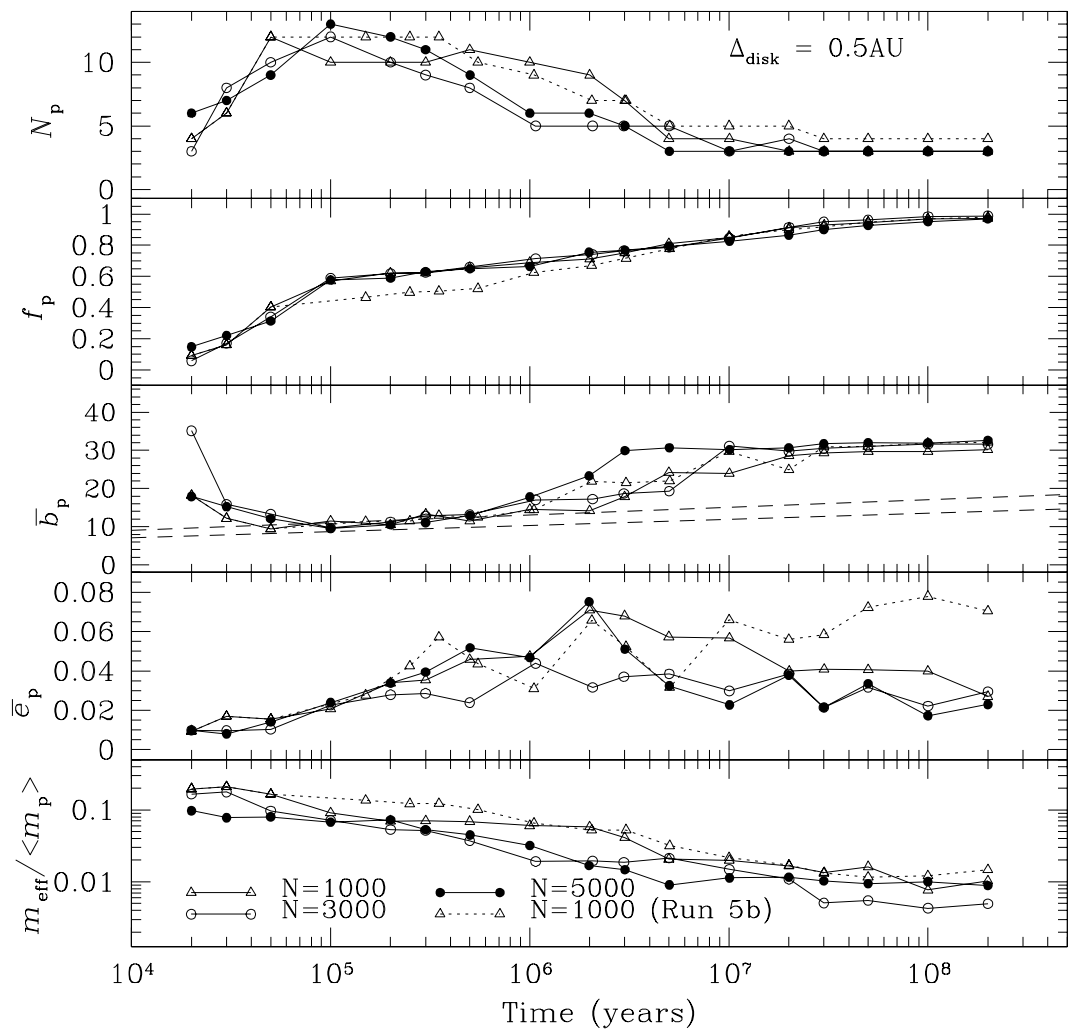}
\end{center}
\caption{Same as Figure~4 but for the case of 
$\Delta_{\rm disk} = 0.5$AU (Runs~5, 5b, 6, and 7).}
\end{figure}

\clearpage
\begin{figure}
\begin{center}
\includegraphics[width=.85\textwidth]{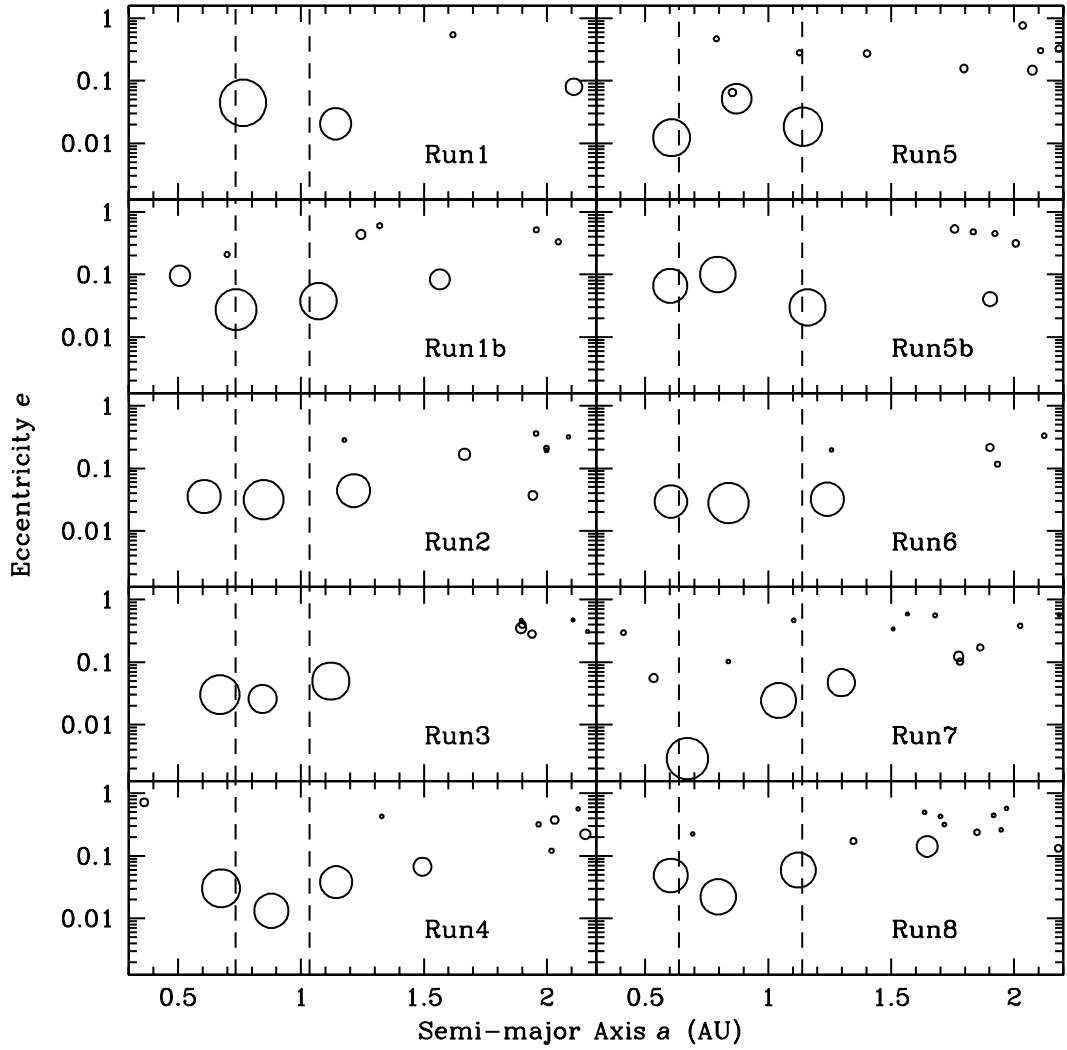}
\end{center}
\caption{Snapshot of all of runs on the $a$--$e$ plane at 200Myr.
The vertical lines represents inner and outer edges of 
the initial planetesimal disk.}
\end{figure}

\clearpage
\begin{figure}
\begin{center}
\includegraphics[width=.7\textwidth]{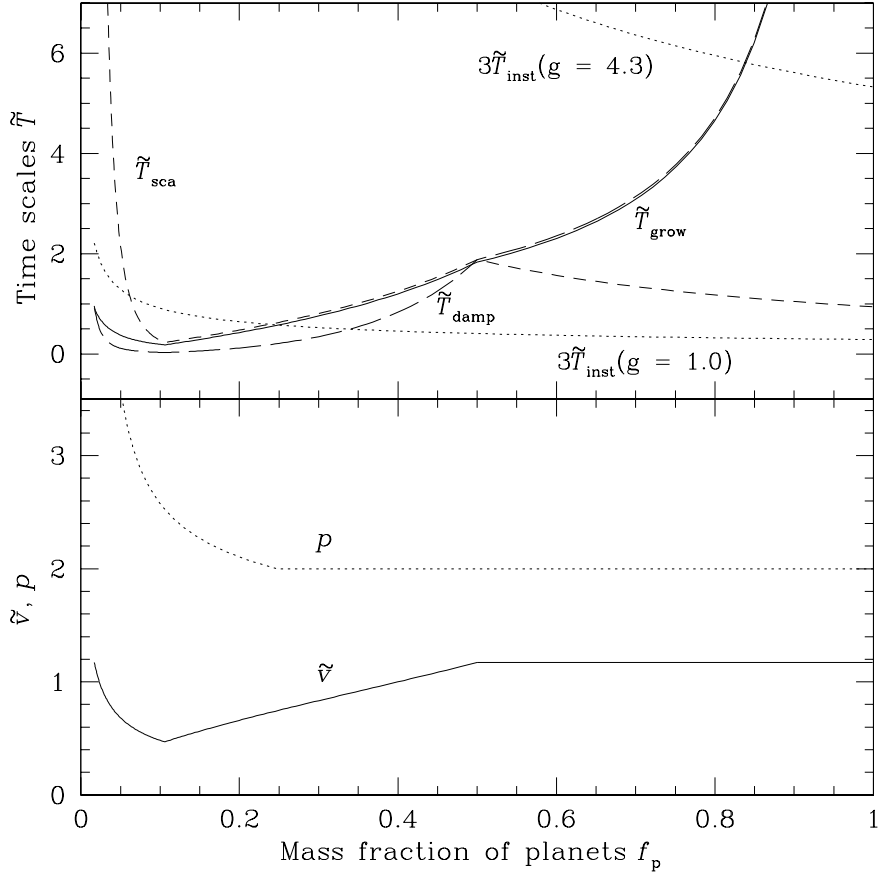}
\end{center}
\caption{Evolution of the normalized time scales $\tilde{T}$ 
(upper panel),  the velocity of planetesimals $\tilde{v}$ 
normalized by the escape velocity of 
planets, and the power-law index, $q$, 
for mass distribution of planetesimals 
(lower panel) as functions of mass fraction of planets $f_{\rm p}$ 
to the total mass.  $\tilde{T}_{\rm grow}$,  $\tilde{T}_{\rm scat}$,
 $\tilde{T}_{\rm damp}$, and $\tilde{T}_{\rm inst}(g)$ represent 
time scales for the growth of planets, evolution of 
the planetesimal velocity due to scattering by planets, 
damping of eccentricities of planets due to the dynamical friction of 
planetesimals, and orbital instability for multiple planet systems,
respectively. $\tilde{T}_{\rm grow}$ is slightly shifted downward (by 0.05) 
to avoid overlapped displays of the time scales.}

\end{figure}

\clearpage
\begin{figure}
\begin{center}
\includegraphics[width=.65\textwidth]{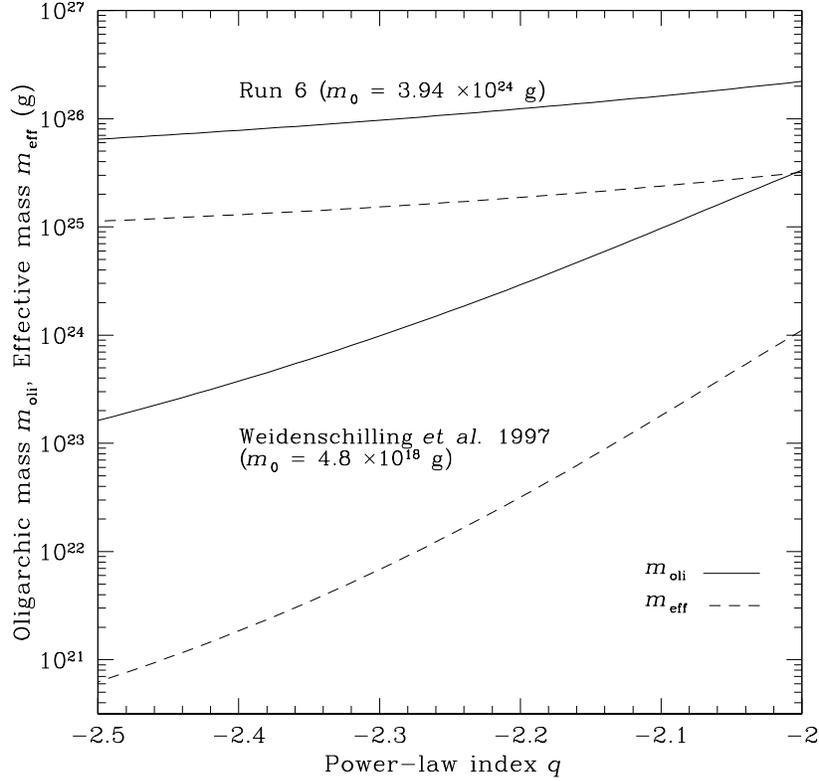}
\end{center}
\caption{The minimum oligarchic mass $m_{\rm p, oli}$ (solid lines)
and the effective mass of planetesimals $m_{\rm eff}$ (dashed lines)
verses the power-law index $q$.
The case for our Run~6
($m_0 = 3.94 \times 10^{24}$g, $\Sigma = 19.1$ g cm$^{-2}$, and 
$a = 0.89$AU), and the case for parameters used in  
simulations of \citet{Wei97}
($m_0 = 4.8 \times 10^{18}$ g, $\Sigma = 16.7$ g cm$^{-2}$, and 
$a = 1.0$AU) are shown. The latter parameters are also used in 
\citet{WS93} and \citet{Ina01}.}

\end{figure}

\clearpage
\begin{figure}
\begin{center}
\includegraphics[width=.65\textwidth]{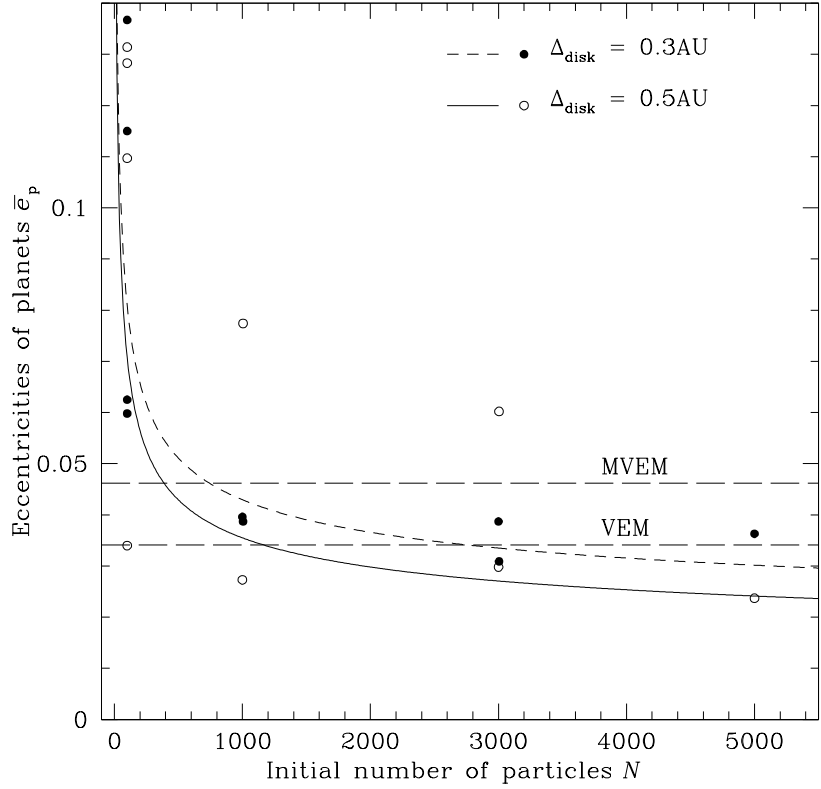}
\end{center}
\caption{Comparison of the final eccentricities of planets between 
$N$-body simulations (open and filled circles; from Table~1 except for $N = 100$) 
and analytic estimates
(solid and dashed lines; eq.~[\ref{eq:eeq2}]). In analytic estimates, we assume 
the mass of planets $m_{\rm p}$ to be $0.66 M_{\rm E}$ and 
$q = -2$ for remnant planetesimals.
MVEM and VEM stand for the values for the 
current terrestrial planets
with and without Mercury, respectively.
}

\end{figure}

\clearpage


\begin{deluxetable}{ccccccc}
\tabletypesize{\footnotesize}
\tablewidth{0pt}
\tablecaption{Initial conditions and final states of simulations}
\tablehead{
\colhead{Run} & \colhead{$\Delta_{\rm disk}$(AU)} & \colhead{$N$} & 
\colhead{$\alpha$} & \colhead{$N_{\rm p}$} & 
\colhead{$\bar{e}_{\rm p}$ ($10^{-2}$)}  & 
\colhead{$S_{\rm d}$ ($10^{-3}$)}}
\startdata
 1 &  0.3 &  1000  & -1 & 3 & 3.96  & 1.55   \\ 
 1b&  0.3 &  1000  & -1 & 4 & 3.87  & 1.51   \\ 
 2 &  0.3 &  3000  & -1 & 3 & 3.87  & 1.79   \\ 
 3 &  0.3 &  5000  & -1 & 3 & 3.63  & 0.99   \\ 
 4 &  0.3 &  3000  & -2 & 4 & 3.09  & 0.99   \\ 
 5 &  0.5 &  1000  & -1 & 3 & 2.73  & 0.95   \\ 
 5b&  0.5 &  1000  & -1 & 4 & 7.74  & 3.49   \\ 
 6 &  0.5 &  3000  & -1 & 3 & 2.98  & 0.60   \\ 
 7 &  0.5 &  5000  & -1 & 3 & 2.37  & 0.49   \\ 
 8 &  0.5 &  3000  & -2 & 4 & 6.02  & 2.97   \\  \tableline
 MVEM    &        &        &   & 4 & 4.62 & 1.90    \\   
 VEM     &        &        &   & 3 & 3.41 & 1.50    \\   
\enddata
\tablecomments{Parameters $\Delta_{\rm disk}$, $N$,  and $\alpha$
stand for the width, number of planetesimals, and 
power-law index for the surface density
of initial planetesimal disks, respectively,  
and $N_{\rm p}$, $\bar{e}_{\rm p}$, and $S_{\rm d}$ 
stand for the number, averaged orbital eccentricity (eq.~[\ref{eq:aveep}]),
and angular momentum 
deficit of planets (eq.~[\ref{eq:sd}]) at the end of simulations.
In Runs~1b and 5b we switch integrators and 
reduce $g$ to unity at $5 \times 10^{4}$ yr, and
at $10^{5}$ yr for other runs. 
MVEM and VEM stand for the current terrestrial planets
with and without Mercury, respectively.} 
\end{deluxetable}

\end{document}